\documentclass[preprint,5p,twocolumn]{elsarticle}
\usepackage{moreverb,url}
\usepackage[colorlinks,bookmarksopen,bookmarksnumbered,citecolor=red,urlcolor=red]{hyperref}

\usepackage[font=footnotesize,caption=false]{subfig}
\usepackage{booktabs}
\usepackage{hyphenat}
\usepackage{multirow}
\usepackage{textcomp}
\usepackage{amsmath}

\begin{document}


\title{Pricing Schemes for Energy-Efficient HPC Systems:\\ 
Design and Exploration}





\author[DISI,DEI]{Andrea~Borghesi}
\ead{andrea.borghesi3@unibo.it}
\author[DEI]{Andrea~Bartolini}
\ead{a.bartolini@unibo.it, barandre@iis.ee.ethz.ch}
\author[DISI]{Michela~Milano}
\ead{michela.milano@unibo.it}
\author[DEI,ISL]{Luca~Benini}	
\ead{luca.benini@unibo.it,luca.benini@iis.ee.ethz.ch}

\address[DISI]{DISI, University of Bologna. Viale Risorgimento 2, 40123, Bologna, Italy}
\address[DEI]{DEI, University of Bologna. Viale Risorgimento 2, 40123, Bologna, Italy}
\address[ISL]{Integrated Systems Laboratory at ETH Zurich, Switzerland}

\newcommand{\timeframe}{\theta}
\newcommand{\totalCores}{NC^T}
\newcommand{\pue}{PUE}
\newcommand{\electricityPrice}{E_{\epsilon}}
\newcommand{\lifetime}{LF}
\newcommand{\systemCostTotal}{C_S^T}
\newcommand{\roi}{ROI}
\newcommand{\energyITYear}{C_{EI}^Y}
\newcommand{\jobDurMF}{\delta_M}
\newcommand{\avgJobCores}{\nu_j}
\newcommand{\scalingFactor}{\varphi}
\newcommand{\scalingFactorFixed}{\phi}
\newcommand{\jobSensitivity}{\sigma}
\newcommand{\utilization}{U}
\newcommand{\alphaFactor}{\alpha}
\newcommand{\idlePerc}{\iota}
\newcommand{\nonSlowPerc}{\beta}

\newcommand{\systemCostYear}{C_S^Y}
\newcommand{\energyCoolYear}{C_{EC}^Y}
\newcommand{\energyITLifetime}{C_{EI}^T}
\newcommand{\energyCoolLifetime}{C_{EC}^T}
\newcommand{\energyTotLifetime}{C_E^T}
\newcommand{\systemCostTimeframe}{C_S^\timeframe}
\newcommand{\systemCost}{C_S}
\newcommand{\coeffTot}{\kappa_T}
\newcommand{\coeffSys}{\kappa_E}
\newcommand{\corePowerMF}{P}
\newcommand{\coreIdlePowerMF}{P^I}
\newcommand{\coreActivePowerMF}{P^A}
\newcommand{\jobPowerMF}{\pi_M}
\newcommand{\jobDurSF}{\delta_S}
\newcommand{\jobPowerSF}{\pi_S}
\newcommand{\njobsTimeframe}{\xi^\theta}
\newcommand{\njobs}{\xi}
\newcommand{\nActiveRes}{R^a}

\newcommand{\incomeTimeframe}{I^\timeframe}
\newcommand{\income}{I}
\newcommand{\energyITTimeframe}{C_{EI}^\timeframe}
\newcommand{\energyIT}{C_{EI}}
\newcommand{\energyCoolTimeframe}{C_{EC}^\timeframe}
\newcommand{\energyCool}{C_{EC}}
\newcommand{\costTimeframe}{C_T^\timeframe}
\newcommand{\cost}{C_T}
\newcommand{\sysGainTimeframe}{\gamma^\timeframe}
\newcommand{\sysGain}{\gamma}
\newcommand{\jobCostTimeframe}{\chi^\timeframe}
\newcommand{\jobCost}{\chi}
\newcommand{\sysGainTimeframeNorm}{\gamma_N^\timeframe}
\newcommand{\sysGainNorm}{\gamma_N}
\newcommand{\jobCostTimeframeNorm}{\chi_N^\timeframe}
\newcommand{\jobCostNorm}{\chi_N}

\newcommand{\modelOne}{Scheme 1}
\newcommand{\modelTwo}{Scheme 2}
\newcommand{\modelThree}{Scheme 3}
\newcommand{\modelFour}{Scheme 4}

\newcommand{\modelOneEM}{\emph{Pricing Scheme 1}}
\newcommand{\modelTwoEM}{\emph{Pricing Scheme 2}}
\newcommand{\modelThreeEM}{\emph{Pricing Scheme 3}}
\newcommand{\modelFourEM}{\emph{Pricing Scheme 4}}

\begin{abstract}
Energy efficiency is of paramount importance for the sustainability of HPC systems. Energy consumption limits the peak performance of supercomputers and accounts for a large share of total cost of ownership. Consequently, system owners and final users have started exploring mechanisms to trade off performance for power consumption, for example through frequency and voltage scaling.

However, only a limited number of studies have been devoted to explore the economic viability of performance scaling solutions and to devise pricing mechanisms fostering a more energy-conscious usage of resources, without adversely impacting return-of-investment on the HPC facility.
We present a parametrized model to analyze the impact of frequency scaling on energy and to assess the potential total cost benefits for the HPC facility and the user. We evaluate four pricing schemes, considering both facility manager and the user perspectives. We then perform a design space exploration considering current and near-future HPC systems and technologies.
\end{abstract}

\begin{keyword}
High Performance Computing, Energy-Efficiency, Power Consumption, Pricing Schemes
\end{keyword}

\maketitle

\section{Introduction}
Energy consumption poses a great challenge for the growth of worldwide HPC installations. As supercomputers increase their peak performance, so do their power consumption, leading in turn to increased energy costs. Hence, the last few years saw a shift from a ``performance at all costs'' mentality to a more balanced and energy efficient perspective \cite{green500,exascaleAcceptedPower,USdatacenter_energy_report}.  

Several methods aim at curtailing the power consumption through a trade-off between the computing performance and power consumption, for example via frequency and/or voltage scaling (\emph{DVFS}) \cite{rountree2007bounding}. The main drawback of this technique is the decreased computing speed that leads to increased application run times. This issue is partially mitigated because several HPC applications and benchmarks are not CPU-bound but present a memory and I/O bottleneck \cite{Zivanovic:2017:MMH:3058793.3023362,marjanovic2014performance,Radulovic:2015:TWM:2818950.2818955}; reducing the frequency of the computing units used by these jobs does not impact severely their time-to-solution (TtS)\cite{Auweter2014}. For instance, memory or I/O bound application are less sensitive to power reduction. See differences between CPU-heavy benchmarks such as HPL\cite{dongarra2003linpack} and the memory bandwidth constrained HPCG\cite{dongarra2013toward}. While in the rest of the paper we will refer explicitly to frequency scaling, our conclusions can also be applied Intel's \emph{Running Average Power Limit} (RAPL) \cite{David:2010:RMP:1840845.1840883}, that does not directly change the computing nodes clock frequency but indirectly does so by enforcing a socket-level power cap. This technique is analogous to DVFS since the power bound leads to increased run times \cite{Inadomi:2015:AMI:2807591.2807638,langer2015}.   

While applications of DVFS in power capped contexts have been widely studied, very little attention has been dedicated to the economic aspect of the frequency scaling. For example, a very common accounting scheme in HPC centers consists in linking the price paid by final users to the time-to-solution of their application multiplied by the requested resources \cite{cineca_accounting}; this scheme is therefore directly affected by techniques altering the applications run time. The rapid depreciation of computing facilities pushes against any policy that stretches job execution time. Moreover, decreasing the computing unit performance leads to lower power consumption, but this does not guarantee lower energy consumption, due to the longer durations.  

In this paper we take steps to address these issues. We introduce a parameterized model representing a HPC system, based 
on a real Top 500 supercomputer on the tier-0 \emph{Fermi} supercomputer, hosted at the CINECA computing center\cite{CINECA}. 
We use the model to understand the economic impact of frequency scaling, from the point of view of both the facility manager (maximizing the overall gain and reducing the total cost of ownership -- also called TCO) and of users (minimizing the costs paid for resource per hour). We present four different pricing schemes and we evaluate their economic viability, given the parameters characterizing the \emph{Fermi} supercomputer and the hosting facility. We consider how DVFS impacts both the energy costs (The electricity cost paid by the facility to operate the IT infrastructure plus the cooling system) and the generated income; we explore mechanisms that can be used to foster a reduction in energy costs while maintaining a profitable condition for both users and owners.  We also extend our parametric analysis considering how the pricing schemes could generate different outcomes with different systems and operating conditions. 

The rest of the paper is organized as follows: Section~\ref{sec:related} provides an overview on the related works in the area of frequency scaling in HPC and a brief discussion on energy-aware pricing schemes found in the data center literatures. Section~\ref{sec:model} describes the parameterized models and evaluates the proposed pricing schemes. Section~\ref{sec:alternative_scenarios} discusses the alternative scenarios and explores the design space. Finally, Section~\ref{sec:conclusion} summarizes the paper and provides the concluding remarks.

\section{Related Works}
\label{sec:related}
In this section we briefly describe the state-of-the-art techniques aiming at energy efficiency (in particular frequency scaling). We then present an overview of the literature regarding pricing schemes found in data centers and targeted at fostering energy efficient solutions.

\subsection{Power/Energy Efficiency}
\label{sec:related_powerEff}
Since the HPC community widely recognizes the need to reduce power consumption in supercomputers, several research avenues have been explored for this purpose. Many techniques have been proposed to bound the power consumption of HPC machines, ranging from Dynamic Voltage and Frequency Scaling (DVFS) \cite{Etinski2012579}, energy proportional systems \cite{varsamopoulos2010energy}, over-provisioning \cite{Patki:2013:EHO:2464996.2465009}, turning off idle resources \cite{4536218}, exploiting components variability \cite{shoukourian2015adviser}. In this paper we are going to focus on frequency scaling and socket-level power capping (RAPL) because they are well-known solutions that have been adopted in several HPC systems \cite{Freeh:2005:UME:1065944.1065967,lim2006adaptive,rountree2007bounding,Rountree:2009:AMD:1542275.1542340,Bailey:2014:ACS:2703009.2706736,Patki:2015:PRM:2749246.2749262}.

Nowadays, many supercomputers employ some form of DVFS\cite{lim2006adaptive, 1559985}, i.e. they exchange processor performance for lower power consumption. With DVFS, a processor can run at one of the supported frequency/voltage pairs lower than the maximum one.
The main issue with DVFS-based approaches is the trade-off between power savings and decrease in performance: reducing the clock frequency clearly increases the TtS of the applications. To overcome this issue, several methods try to apply DVFS only in periods of low system activities or in particular phases of a job execution. For example, in \cite{Freeh:2007:AET:1263127.1263246}, Freeh et al. study the energy-time trade-off of high performance cluster nodes with several power states available. They conclude that applying DVFS to applications with memory or communication bottlenecks does not imply large time penalties. This strategy strongly relies on the nature of the running applications, which must be known and modeled in advance, before their actual execution. In \cite{hsu2005power}, Hsu et al. propose to solve this problem through a power-aware \emph{adaptive} algorithm which does not employ any application-specific information a priori, but implicitly gathers such information at run-time. 

Etinski et al. \cite{Etinski2010} extend the well-known EASY-backfilling scheduling policy to limit a supercomputer power consumption through DVFS. Their results are promising in terms of energy savings and also a better utilization of the system and reduced waiting time for the users, thanks to the possibility to execute more jobs concurrently if their frequency (thus power) is reduced. The same authors introduce also another approach in \cite{Etinski2012615}: in the latter work they propose a novel scheduling policy based on integer linear programming (ILP). This method offers better performance in terms of average job wait time over various power budget. These two works focus exclusively on the effect of frequency scaling on applications run times while we are mainly concerned with the energy consumption and its economic impact.

RAPL provides a software configurable and hardware enforced power cap. Instead of setting a specific frequency, this mechanism takes as input the power budget for a socket and subsequently forces the power consumption to be within the limit. For instance, Ellsworth et al. \cite{Ellsworth:2015:DPS:2807591.2807643} present a scheme to decide the power allocated to each node in a supercomputer (\emph{Dynamic Power Sharing}). Initially the overall available power budget is uniformly divided among all nodes; periodically the algorithm adjusts the allocated power depending on actual consumptions, i.e. if a node consumes less power than the allocated one the exceeding capacity can be transferred to a different node which needs it. RAPL is used to enforce the node power limit at run time. The main drawback of RAPL is the same that troubles DVFS mechanisms, namely the indiscriminate power reduction implies an increase in TtS (performance loss). 

The main limitation of the related works in the research literature is that they focus (almost) exclusively on the energy-savings and time-to-solution considerations while discounting the cost aspects. All the considered approaches can influence the HPC system revenues exclusively through the reduction of energy/power spending and therefore overlook a critical component of the facility costs, the non negligible depreciation costs. In our paper we consider both elements that determine the supercomputer TCO. 

Real HPC applications have different sensitivities towards frequency \& voltage scaling; memory or I/O bound application are less sensitive to frequency reduction. For instance, see differences between CPU-heavy benchmarks such as HPL\cite{dongarra2003linpack} and the memory bandwidth constrained HPCG\cite{dongarra2013toward}. We consider for simplicity an ``average'' job sensitivity and sweep it as a parameter.

\subsection{Pricing Schemes}
\label{sec:related_priceSchemes}
Another important area of research deals with the problem of finding optimal pricing schemes for the resources composing a supercomputer. The current state-of-the-art for pricing schemes in HPC systems is somewhat lacking, whilst researchers in the data center community investigated this issue in a more thorough manner\cite{samadi2010optimal,sharma2012pricing,zhao2014dynamic}. Generally speaking, data centers operate with a slightly different set of assumption w.r.t. HPC facilities and therefore they are not directly comparable to the method proposed in this paper.

Chase et al. \cite{chase2001managing} present a new architecture to manage resources in a data center, with the goal of energy efficiency. The main idea is to implement a bidding mechanism where the services running in the system bid for resources as a function of  delivered  performance. Afterwards, resource prices are regulated through a greedy algorithm to balance supply and demand, allocating resources to their most efficient use.

Zhang et al. \cite{zhang2012electricity} consider the issue of minimizing the electricity bill of a network of data centers; for this purpose they devise an approach that leverage the different electricity prices in different geographical locations to distribute workloads among those locations. Their work explicitly models the effects of the power demands induced by cloud-scale data centers on electricity prices and the power consumption of cooling and networking in the minimization of electricity bill. Although the proposed solution is very interesting, the vast majority of nowadays HPC systems do not have a distributed nature similar to the one considered in this work.

Wang et al. \cite{wang2014hierarchical} tackle the problem of optimizing data center electric utility bill under uncertainty in workloads and real-world pricing schemes. They consider a data center where the power consumption of the IT equipment can be modulated via control knobs. The key assumption of the model they propose is that the power effects of most IT control knobs can be seen as dropping and/or delaying a portion of the power demand, i.e. through dynamically modulating the workload. They propose a hierarchical infrastructure to manage system resources and workload; the hierarchical structure allows to separate the abstract layer specifying the optimization policy from the lower level that implements the actual power-modulation knobs. The main drawback back of this work (and several others found in the literature) is that it disregards the total cost of ownership and the depreciation costs.

\section{The HPC System Model}
\label{sec:model}

In this section we introduce the parameterized model, used to describe the cost, energy, performance trade-off in a generic supercomputer. The parameters configuration considered in this section is based on the \emph{Fermi} supercomputer\cite{fermi_supercomputer}. The proposed model abstracts the ensemble of computing resources as a composition of allocable elements. As the considered system was composed of multi-cores, we referred to them as ``core''. This is done to simplify the analysis but nothing prevents the addition of different resources to the model to extend our approach. In our analysis we do assume that scheduling and allocation decisions have been taken by a higher-level scheduler. This is normal in supercomputer infrastructures\cite{Feitelson1997,Feitelson2005,caoCluster06,workloadsCharModel}. 

We assume that the considered machine is capable of decreasing the power consumption of computing units in exchange for reduced performance through frequency \& voltage scaling, which may lead to an increased run-time of the involved applications, accordingly to their properties. We model the power consumption of each computing resource with two contributions: the idle power and the active power. The idle power is a constant power term needed to keep the resource on, the active power is only consumed when the resource is active and executes a job. The absolute value is proportional to the clock frequency. The dependency of the active power to the frequency is monotonic and superlinear with an exponent alpha dependent on the technology\cite{sakurai1990alpha}.

\subsection{Model Description}
\label{sec:model_description}

The key parameters composing the model are listed in Table~\ref{tab:base_parameters}. From these base parameters we compute the values of a set of intermediate variables, presented in Table~\ref{tab:derived_parameters}. In Table~\ref{tab:output_parameters} we report the output, or target, parameters. The main output parameters are used to evaluate the pricing schemes discussed in Section~\ref{sec:model_pricing_schemes}. We chose two main parameters: 1) the system gain (the difference between the income obtained by the system owner and the operating costs); 2) the price paid by users for their application (measured as the price paid per hour and per single resource usage). These outcomes are relative to the considered time frame $\timeframe$. The model parameters are linked through the mathematical expressions exposed in the tables. Some parameters are self explaining; we give here details to illustrate the less obvious ones. 
Part of the parameters presented in Table~\ref{tab:base_parameters} describe the HPC facility. In our case, their values depend on the supercomputer we took as example; different configurations can model different systems. Other parameters are instead used to represent the applications.

There are two main parameters that define the behaviour of the system (how system gain and job price are affected) when frequency scaling is applied:
\begin{itemize}
\item the scaling factor, $\scalingFactor$, indicates how much power consumptions are decreased (the same factor is applied to each slowed down jobs);
\item the job sensitivity, $\jobSensitivity$, modulates the duration increase due to the power scaling (again, same value applied to all slowed jobs)
\end{itemize}

\begin{table*}[bt]
\small\sf\centering
\begin{tabular}{lcr}
 \toprule
 \nohyphens{\emph{Name}} & \qquad \nohyphens{\emph{Symbol}} & \qquad \nohyphens{\emph{Unit}} \\
  \midrule
   Time frame & \qquad $\timeframe$ & \qquad Days \\
   Number of cores in the system & \qquad $\totalCores$ & \qquad NA \\ 
   Power Usage Efficiency & \qquad $\pue$ & \qquad NA \\
   Electricity cost & \qquad $\electricityPrice$ & \qquad \texteuro / KWh \\
   System lifetime & \qquad $\lifetime$  & \qquad Years \\
   System installation cost & \qquad $\systemCostTotal$ & \qquad \texteuro \\
   Estimated energy cost (IT) per year & \qquad $\energyITYear$ & \qquad \texteuro \\
   Return On Investment ($\geq 1$) & \qquad $\roi$ & \qquad NA \\ 
   Percentage of system utilization & \qquad $\utilization$ & \qquad NA \\
   Idle power as \% of power at max. frequency  & \qquad $\idlePerc$ & \qquad NA \\
   Alpha factor & \qquad $\alphaFactor$ & \qquad NA \\ 
   \midrule
   Job TtS (Time-to-Solution) at maximum frequency (estimate) & \qquad $\jobDurMF$ & \qquad Hours \\
   Number of requested cores per job & \qquad $\avgJobCores$ & \qquad NA \\ 
   Frequency scaling factor & \qquad $\scalingFactor$ & \qquad NA \\
   Job sensitivity & \qquad $\jobSensitivity$ & \qquad NA \\
   Percentage of non-slowed jobs & \qquad $\nonSlowPerc$ & \qquad NA \\
  \bottomrule
\end{tabular}
\caption{Model Base Parameters}
\label{tab:base_parameters}	
\end{table*}

The system might be not fully used (not enough jobs, resource bottlenecks, SLAs constraints..) but its cores are occupied only up to a certain percentage $\utilization$. 
In the proposed model we consider also the case where the power consumption is scaled down only for a fraction of the jobs;
$\nonSlowPerc$ tells the percentage of jobs that are not subject to slow down (conversely $1 - \nonSlowPerc$ indicates the fraction of jobs with a reduced power consumption). The alpha factor $\alphaFactor$ is a technology-dependent parameter and affects the reduction in power consumption following a frequency reduction. Given a core base power consumption at maximum frequency, the idle power percentage $\idlePerc$ indicates the proportion of consumption due to the idle power (when the core is not used). Lower values of $\idlePerc$ indicate a more energy proportional system, i.e. systems where power tends to near-zero values when frequency tends to zero.

The scaling factor, $\scalingFactor$, specifies the ratio between the maximum and reduced frequency and it directly modulates the power consumption variation (decrease). It is a real number and, given a maximum frequency $f_{max}$ and the scaled one $f_{scaled}$, is computed as $\scalingFactor = f_{max}/f_{scaled}$. The job sensitivity, $\jobSensitivity$, modulates the time-to-solution increase due to the power scaling. The job sensitivity embeds both the nature of the application (ranging between CPU-bound or memory-bound) and the fact that a HPC job can be composed by several sub-tasks with relative dependencies: an application with many intertwined tasks may experience higher performance degradation when subject to frequency scaling.

\begin{table*}[bt]
\small\sf\centering
\begin{tabular}{lllr}
 \toprule
 \nohyphens{\emph{Name}} & \qquad \nohyphens{\emph{Symbol}} & \qquad \nohyphens{\emph{Expression}}  & \qquad \nohyphens{\emph{Unit}}  \\
  \midrule
   System cost per year (depreciation) & \qquad $\systemCostYear$ & \qquad $\systemCostTotal / \lifetime$ &  \qquad \texteuro \\
   Cooling Energy Cost per Year & \qquad $\energyCoolYear$ & \qquad $\energyITYear \cdot (\pue -1) $ & \qquad \texteuro \\
   IT energy cost - Lifetime & \qquad $\energyITLifetime$ & \qquad $\energyITYear \cdot \lifetime$ & \qquad \texteuro \\
   Cooling energy cost - Lifetime & \qquad $\energyCoolLifetime$ & \qquad $\energyCoolYear \cdot \lifetime$ & \qquad \texteuro \\
   Total energy cost - Lifetime & \qquad $\energyTotLifetime$ & \qquad $\energyITLifetime + \energyCoolLifetime$ & \qquad \texteuro \\
   System cost (depreciation) - Time frame & \qquad $\systemCostTimeframe$ & \qquad $\systemCostYear / 365 \cdot \timeframe $ &  \qquad \texteuro \\
   Coefficient - Total &  \qquad $\coeffTot$ & \qquad $\frac{\roi \cdot \systemCostTotal + \energyTotLifetime}{\totalCores \cdot \lifetime \cdot 24\cdot365}$ & \qquad NA\\ 
   Coefficient - System Only &  \qquad $\coeffSys$ & \qquad $\frac{\roi \cdot \systemCostTotal}{\totalCores \cdot \lifetime \cdot 24\cdot365}$ & \qquad NA\\ 
\midrule
   Core Power (max frequency) & \qquad \corePowerMF & \qquad $ \frac{1000 \cdot \energyITYear / \electricityPrice}{\totalCores\cdot 365\cdot 24}$ & \qquad W \\
   Core Idle Power (max frequency) & \qquad $\coreIdlePowerMF$ & \qquad $\idlePerc \corePowerMF $ & \qquad W \\
   Core Active Power (max frequency) & \qquad $\coreActivePowerMF$ & \qquad $(1-\idlePerc) \cdot \corePowerMF$ & \qquad W \\
   Job power consumption at max frequency & \qquad $\jobPowerMF$ & \qquad $\avgJobCores (\coreIdlePowerMF+\coreActivePowerMF)$ & \qquad W  \\
\midrule
   Job TtS at scaled frequency & \qquad $\jobDurSF$ & \qquad $\jobDurMF + \jobDurMF (\scalingFactor-1) \jobSensitivity$ & \qquad Hours \\
   Job power consumption at scaled frequency & \qquad $\jobPowerSF$ & \qquad $\avgJobCores (\coreIdlePowerMF +\frac{\coreActivePowerMF}{\scalingFactor^\alphaFactor})$ & \qquad W  \\
\midrule
   Number of resources active & \qquad $\nActiveRes$ & \qquad $\totalCores \times \utilization$ & \qquad \# Cores \\
  \bottomrule
\end{tabular}
\caption{Model Derived Parameters}
\label{tab:derived_parameters}	
\end{table*}

The idle and active power consumed by each core at maximum frequency ($\coreIdlePowerMF$ and $\coreActivePowerMF$) are obtained by dividing the total energy consumed by the IT infrastructure -- derived from the yearly energy cost $\energyITYear$ -- by the total number of core and the hours of utilization. 
The power consumption of a job at maximum frequency is computed as the sum of idle and power consumption for each core (at maximum frequency) multiplied by the number of requested cores ($\avgJobCores$). In Table~\ref{tab:derived_parameters} we also observe how the time-to-solution and the power consumption of a job change if the power is scaled down; the scaling factor $\scalingFactor$ and the job sensitivity $\jobSensitivity$ are the only parameters affecting the outcome -- we assume that $\alphaFactor$ remains constant in the whole time frame (besides being the same for all cores). 


The parameter $\nActiveRes$ indicates the number of resources (only cores in our model) that are used in the system by the running applications; it is computed as the number of total resources available in the system multiplied by the system utilization $\utilization$.

\begin{table*}[bt]
\small\sf\centering
\begin{tabular}{lclr}
 \toprule
 \nohyphens{\emph{Name}} & \qquad \nohyphens{\emph{Symbol}} & \qquad \nohyphens{\emph{Expression}}  & \qquad \nohyphens{\emph{Unit}}  \\
  \midrule
   Income per time frame & \qquad $\incomeTimeframe$ & \qquad Sec.~\ref{sec:model_pricing_schemes} & \qquad \texteuro \\
   IT energy cost per time frame & \qquad $\energyITTimeframe$ & \qquad $[(1-\nonSlowPerc)\jobPowerSF\jobDurSF+\nonSlowPerc\jobPowerMF\jobDurMF] \cdot \frac{\electricityPrice}{1000}$ & \qquad \texteuro \\
   Cooling energy cost per time frame & \qquad $\energyCoolTimeframe$ & \qquad $\energyITTimeframe \cdot (\pue - 1) $ & \qquad \texteuro \\
   Total cost per time frame & \qquad $\costTimeframe$ & \qquad $\systemCostTimeframe + \energyITTimeframe + \energyCoolTimeframe$ & \qquad \texteuro \\
   \midrule
   System gain per time frame & \qquad $\sysGainTimeframe$ & \qquad $\incomeTimeframe - \costTimeframe$ & \qquad \texteuro \\
   Average job price & \qquad $\jobCostTimeframe$ & \qquad $\frac{\incomeTimeframe\avgJobCores}{\nActiveRes}$  & \qquad \texteuro \\
  \bottomrule
\end{tabular}
\caption{Model Output Parameters}
\label{tab:output_parameters}	
\end{table*}

The table contains also the derived parameters which are directly involved in the computation of the final output variables, in particular the total cost, depreciation payment plus energy consumption, per time frame. We assume that the depreciation cost is constant in the time period $\systemCostTimeframe$, the energy cost for the cooling is proportional to the IT energy cost, the latter being the sum of the energy consumption of each job. As discussed earlier, only a percentage of jobs undergo a slow down, therefore the energy consumption of each job is a combination of TtS $\times$ power at maximum frequency (non-slowed down jobs) and TtS $\times$ power at scaled frequency (slowed down jobs). The sum of all job energies is multiplied by the electricity cost to infer the energy costs ($\electricityPrice/1000$). We assume that the energy costs are going to be identical for each pricing scheme presented in Section~\ref{sec:model_pricing_schemes} (the pricing scheme influences only the system income and not its expenses). 

The $\roi$ is an input parameter representing the expected Return-On-Investment desired by the system owner. $\coeffTot$ stands for the baseline hourly cost per resources, derived from $\roi$, depreciation and estimated energy cost. $\coeffSys$ is defined similarly but discarding the energy cost. The maintenance costs and the value of money are embedded in the depreciation costs and Return-Of-Investment.


\subsection{Energy Saving Potential}
\label{sec:model_energy_savings}

A fundamental aspect impacting the system cost -- hence system gain and price paid by users -- is the energy cost. We must consider two issues: 1) does decreasing speed (clock frequency) actually reduce energy consumption? if that is the case, certainly the energy cost would go down; 2) even if the above is true, does this lead to reduced system TCO? This may not happen because of depreciation costs. In this section we are going to answer to the first question, while Sec.~\ref{sec:model_pricing_schemes} deals with the second issue.

In general, when we decrease the power consumption of a set of computational resources the HPC jobs that are using them will suffer a performance loss and thus they might require more time to complete. The power decrease and time-to-solution increase are clearly intertwined and their relation strongly depends on the nature of the application; for instance, a memory-bound application would experience a smaller TtS increase. This may lead to an actual energy consumption increase since the energy $E$ associated to a job is computed as: $E = \pi \times \delta$, where $\pi$ is the power consumption of the job and $\delta$ is its time-to-solution.

To answer the question we can analyze the ratio between the energy consumed by a job at maximum frequency and the energy consumed at the reduced frequency. The energy ratio value is expressed by the following equation:
\begin{align}
E_{ratio} & = \frac{\jobPowerMF \times \jobDurMF}{\jobPowerSF \times \jobDurSF} = 
\frac{\avgJobCores (\coreIdlePowerMF+\coreActivePowerMF) \times \jobDurMF}
{\avgJobCores (\coreIdlePowerMF +\frac{\coreActivePowerMF}{\scalingFactor^\alphaFactor} ) \times (\jobDurMF + \jobDurMF (\scalingFactor-1) \jobSensitivity)} \nonumber \\
& = \frac{(\idlePerc \corePowerMF+(1-\idlePerc) \cdot \corePowerMF) \times \jobDurMF}
{(\idlePerc \corePowerMF +\frac{(1-\idlePerc) \cdot \corePowerMF}{\scalingFactor^\alphaFactor} ) \times (\jobDurMF + \jobDurMF (\scalingFactor-1) \jobSensitivity)} \nonumber \\
& = \frac{1}{(\idlePerc +\frac{(1-\idlePerc)}{\scalingFactor^\alphaFactor} ) \times (1 + (\scalingFactor-1) \jobSensitivity)}
\label{eq:energy_ratio}
\end{align}
The numerator and the denominator represent, respectively, the energy consumed by an application at maximum frequency (TtS multiplied by power, $\jobPowerMF \times \jobDurMF$) and the energy consumed at the reduced frequency ($\jobPowerSF \times \jobDurSF$). The rest of the equation is obtained by substituting the TtS and power values with their corresponding expressions, as described in the Tables~\ref{tab:base_parameters}~and~\ref{tab:derived_parameters}. We assume that the parameters that do not appear in Eq.~\ref{eq:energy_ratio} have fixed values.

We observe two facts: 1) values $\geq 1$ are better since they imply that the energy of the job decreases when we scale down its power; 2) the only involved variables are the alpha factor $\alphaFactor$, the idle power expressed as percentage of the total core power (at maximum frequency) $\idlePerc$, the scaling factor $\scalingFactor$ and the job sensitivity $\jobSensitivity$. To further simplify our analysis we now assume that the scaling factor is fixed to a particular value $\scalingFactor> 1$ (it must be greater than one if we want to study the power savings effect); as we are going to see, setting the scaling factor to a constant value does not invalidate our conclusions.

\begin{figure}
\centering
\includegraphics[width=.5\textwidth]{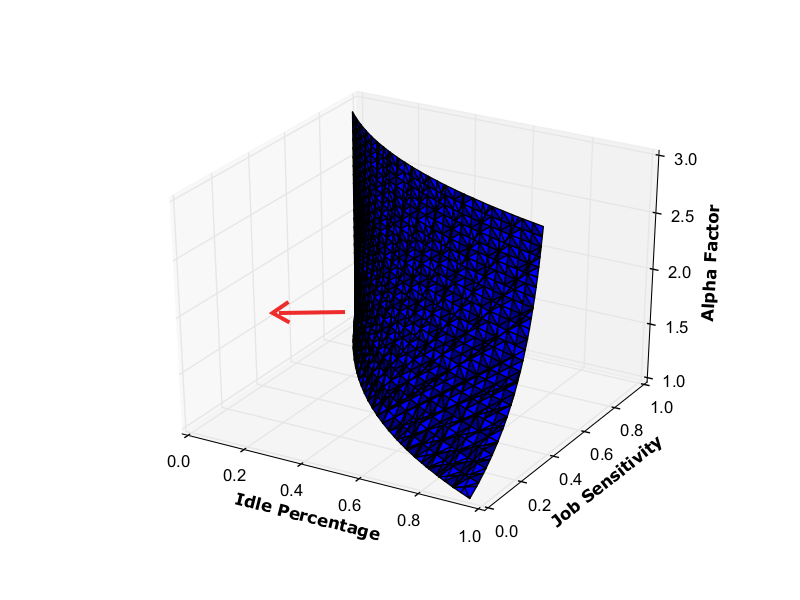}
\caption{Energy Savings: isosurface with energy ratio = 1}
\label{fig:Energy_Ratio_Surface_energy_ratio_surface}
\end{figure}

In Figure~\ref{fig:Energy_Ratio_Surface_energy_ratio_surface} we have a three-dimensional plot representing the isosurface of value 1 corresponding to the energy ratio described in Eq.~\ref{eq:energy_ratio}. The $x$-axis, $y$-axis and $z$-axis contain, respectively, the idle power percentage $\idlePerc$, the the job sensitivity $\jobSensitivity$ and the alpha factor $\alphaFactor$. An \emph{isosurface} is a surface that represents points of constant target value (it is the 3-d analog of an isoline or contour line); points above the isosurface have values larger than the target one, points below the surface have value smaller than the target. The red arrow indicates the volume of space formed by points above the isosurface. For example, in the graph of Fig.~\ref{fig:Energy_Ratio_Surface_energy_ratio_surface} the point with coordinates $(0.2,0.2,2.0)$ is above the isosurface, hence its corresponding energy ratio is larger than 1, which basically means that reducing the clock speed is convenient energy-wise; conversely, the point with coordinates $(0.8,0.8,1.5)$ is situated below the isosurface and corresponds to an energy ratio lower than 1. 

Figure~\ref{fig:Energy_Ratio_Surface_energy_ratio_surface} reveals that indeed there are some combination of values for which reducing the job's power consumption leads to energy savings. For example, as one could have expected, low values of job sensitivity imply a larger energy ratio (saving): if the job TtS increases only marginally when the power is reduced the outcome is an energy saving. We can also notice that better (higher) energy ratios are associated to lower values of $\idlePerc$: this happens because if the idle power component has a relatively smaller influence, decreasing the operating frequency of the computing nodes leads to greater power savings -- the idle power consumption is not affected by the scaling-down action. However, it is also clear that there are many configurations where frequency (and power) reduction does not reduce energy. 

As a first result of the proposed model: cost reduction policies based on performance scaling make sense only if the system is operated in the area above the isosurface, defined by $(\idlePerc, \jobSensitivity, \alphaFactor)$. $\jobSensitivity$ depends on the application slack which is defined based on the target architecture and applications set. $\idlePerc$ and $\alphaFactor$ are instead technological parameters: $\alphaFactor$ is determined by the technology while $\idlePerc$ depends on the system architecture and on the leaking components present in the compute node (i.e. Fans, HDDs, NIC, etc).

\begin{figure}
\centering
\includegraphics[width=0.5\textwidth]{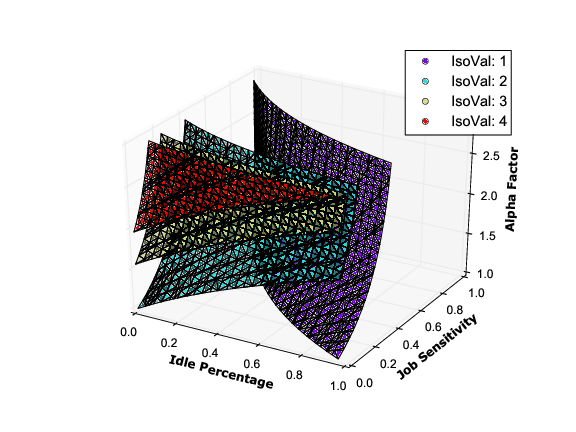}
\caption{Energy Savings: multiple isosurfaces with different ratio values}
\label{fig:Energy_Ratio_Surface_Multi_energy_ratio_surface_multi}
\end{figure}

In Figure~\ref{fig:Energy_Ratio_Surface_Multi_energy_ratio_surface_multi} we displayed different isosurfaces along with the one corresponding to an energy ratio of $1$. The additional isosurfaces correspond to energy ratios of $2$, $3$ and $4$; as noted before, a higher energy ratio means more potential energy saving and thus combinations of $(\idlePerc, \jobSensitivity, \alphaFactor)$ leading towards the new isosurfaces are preferable. 

\begin{figure}
\centering
\includegraphics[width=.5\textwidth]{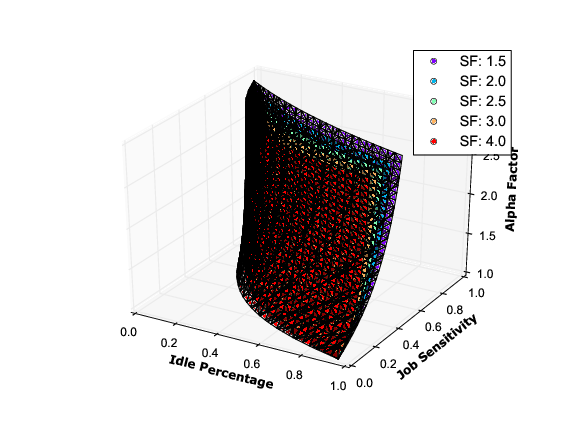}
\caption{Energy Savings: isosurfaces with ratio = 1, different scaling factor $\scalingFactor$ values}
\label{fig:Energy_Ratio_Surface_MultiSF_energy_ratio_surface_multi_sf}
\end{figure}

Figure~\ref{fig:Energy_Ratio_Surface_MultiSF_energy_ratio_surface_multi_sf} shows what happens if we also change the value of the scaling factor parameter $\scalingFactor$; the figure presents again isosurfaces of value $1$. As we anticipated before, the scaling factor influences the energy ratio as revealed by the different gradients of the surfaces but the overall shape of the isosurfaces remain similar. One thing that can be noticed is that when the scaling factor increases the alpha factor impact slightly decreases -- the surface varies less along the $z$-axis.

From Fig.~\ref{fig:Energy_Ratio_Surface_energy_ratio_surface},~\ref{fig:Energy_Ratio_Surface_Multi_energy_ratio_surface_multi}~and~\ref{fig:Energy_Ratio_Surface_MultiSF_energy_ratio_surface_multi_sf} 
we can draw a positive conclusion. Reducing the power consumption of the application in a HPC system can lead to energy savings, depending on some the parameters characterizing the system and the application. 
As a general rule, we can say that facility owner as well as user should target the reduction of power consumption of the less sensitive jobs, i.e. those jobs whose time-to-solution will not be too affected by the power reduction (for example memory, I/O and communication bound applications). 
This conclusion is more prominent in installations in which the idle power is a large component of the total power consumption; In this case reducing the operational frequency can increase the consumed energy.

\subsection{Pricing Schemes}
\label{sec:model_pricing_schemes}
The results of the previous section suggest that, depending on the characteristics of applications (jobs) and of the supercomputer infrastructure, it is possible to decrease the energy consumption of the HPC system by slowing it down. We have not determined yet: 1) if the energy reduction leads to lower costs for the facility manager and for the users; 2) how to perform accounting in order to foster the adoption (by the facility manager and users) of the energy-efficient operating condition. 

We will now discuss four different pricing scheme to see how they impact the TCO and the system total gain and the average job price. In addition to the variables introduced in Table~\ref{tab:output_parameters}, we are also going to consider normalized values for the two most interesting variables: 1) normalized system gain $\sysGainTimeframeNorm$ and 2) normalized job price $\jobCostTimeframeNorm$. The normalized gains and costs are computed w.r.t. to the \modelOneEM~(see \ref{sec:scheme1}), with a scaling factor equal to 1, a situation that we assume is our baseline. The normalized gain (cost) for any given combination of parameters and pricing scheme is obtained by dividing the resulting gain (cost) by the baseline gain (cost). 
Since in all the remaining discussion we are going to focus on system gains and average job prices (and related parameters) computed in time frame $\timeframe$ we are going to omit the time frame reference from the mathematical notation, for the sake of clarity (for example $\sysGainTimeframe \rightarrow \sysGain$). 

In Table~\ref{tab:price_schemes_incomes} the different ways to compute the system time frame income implied by the different pricing schemes are summarized. The table final three columns serve to quickly summarize the scheme features. \emph{Coeff.} indicates the cost coefficient used to give a price to resource per hour; it can include both the depreciation costs (derived from the system installation cost) and the energy cost (``Depreciation+Energy'') or consider only the depreciation cost (``Depreciation''). The \emph{TtS} column specifies the time-to-solution used in the price formula; allowed values are: the real TtS, the oracle TtS (the time-to-solution at maximum frequency) and the scaled time-to-solution (the real TtS divided by the scaling factor).

Finally, the \emph{Energy} columns tells how the energy is taken into account; ``explicit'' means that the energy costs is directly covered by the users, ``implicit'' means that the cost is included in the price coefficient (see the numerator of $\coeffTot$ in Table~\ref{tab:derived_parameters}). 

\begin{table*}[bt]
\small\sf\centering
\begin{tabular}{lllll}
 \toprule
 \nohyphens{\emph{Scheme}} & \qquad \nohyphens{\emph{Expression}} & \qquad \nohyphens{\emph{Coeff.}} & \qquad \nohyphens{\emph{TtS}} & \qquad \nohyphens{\emph{Energy}}  \\
  \midrule
   \modelOne & \qquad $\coeffTot\nActiveRes\big((1-\nonSlowPerc)\jobDurSF+\nonSlowPerc\jobDurMF\big)$ & \qquad Depreciation + Energy & \qquad Real & \qquad Implicit \\ 
   \modelTwo & \qquad $\coeffTot\nActiveRes\big(\jobDurMF\big)$  & \qquad Depreciation + Energy & \qquad Oracle & \qquad Implicit  \\ 
   \modelThree & \qquad $\coeffTot\nActiveRes\big(\frac{(1-\nonSlowPerc)\jobDurSF}{\scalingFactor}+\nonSlowPerc\jobDurMF\big)$  & \qquad Depreciation + Energy & \qquad Scaled & \qquad Implicit  \\ 
	\modelFour \qquad & $\nActiveRes\coeffSys\big((1-\nonSlowPerc)\jobDurSF+\nonSlowPerc\jobDurMF\big)+$ & \qquad Depreciation & \qquad Real & \qquad Explicit \\
   & + $\big((1-\nonSlowPerc)\jobPowerSF\jobDurSF+\nonSlowPerc\jobPowerMF\jobDurMF\big) \cdot \frac{\electricityPrice}{1000}$ \\
  \bottomrule
\end{tabular}
\caption{Income functions with different pricing strategies}
\label{tab:price_schemes_incomes}	
\end{table*}

Since we are interested in understanding the influence of frequency scaling, we begin by focusing our analysis on the parameters that mostly impact its effect, namely the scaling factor ($\scalingFactor$) and job sensitivity ($\jobSensitivity$).

We then observe the target output as a function of these two variables, keeping all remaining parameters fixed. The scaling factor is the main variable the system manager and the users can use as a knob to regulate the power consumption; in our analysis we consider values ranging from 1 (no scaling) to 5 (aggressive power reduction). As an example in today high end CPUs it is common to see the clock frequency ranging from 3.6 GHz (Turbo mode) to 1.2GHz. The job sensitivity has a big influence on the outcome due to the direct impact on the job time-to-solution when the power is reduced; we let the job sensitivity vary from 0, that is an idealized case where reducing the power consumption does not entail a TtS increase, to 1, when the TtS increase is proportional to the power reduction. Job sensitivity values closer to 0 represent memory or I/O bound jobs while moving closer to the opposite end of the range the application are getting more CPU-bound.

When looking at the normalized system gain values larger than one indicate that the considered price model with the specified scaling factor and job sensitivity (tuple $<price\_model, \scalingFactor, \jobSensitivity>$) leads to larger gains w.r.t. to the baseline. Conversely, normalized system gains smaller than 1 and negative values indicate that the baseline produces better results; negative values are possible because for some pricing scheme and parameters combination the system gain can actually be negative -- the system is losing money due to the fact that the cost is higher than the income. With the fixed parameters configuration used in the following subsections the baseline does produce positive net gain for the system. The same discussion can be applied to the normalized job price, with the exception that the latter can never be negative -- the minimum value for the average cost of a job is zero.

One last point to address before introducing the pricing schemes is the issue of the TtS increase. Users might not accept the fact that the TtS of their application is stretched over a certain point due to the frequency scaling. This is mitigated by the fact that when users submit their job, they typically provide estimated TtS that are longer than the actual TtS; stretching their application but maintaining them under their estimated TtS would generate no complaints. Using historical data from a tier-0 supercomputer we discovered that the average ratio between estimated TtS and real TtS is $1.5$ (considering only jobs which run longer than 1 hour to exclude very short application that would skew the mean value). 
This acceptable TtS increase corresponds to the values of scaling factor $\scalingFactor$ and job sensitivity $\jobSensitivity$ displayed as dashed black lines in the following two-dimensional figures and as a black line in the three-dimensional ones. Points below the line correspond to acceptable TtS increase. This information can be used while devising pricing scheme in order to account also for the user satisfaction (for instance, not selecting scaling factor values that would exceedingly slow down an application).

\begin{figure}
\centering
\includegraphics[width=0.5\textwidth]{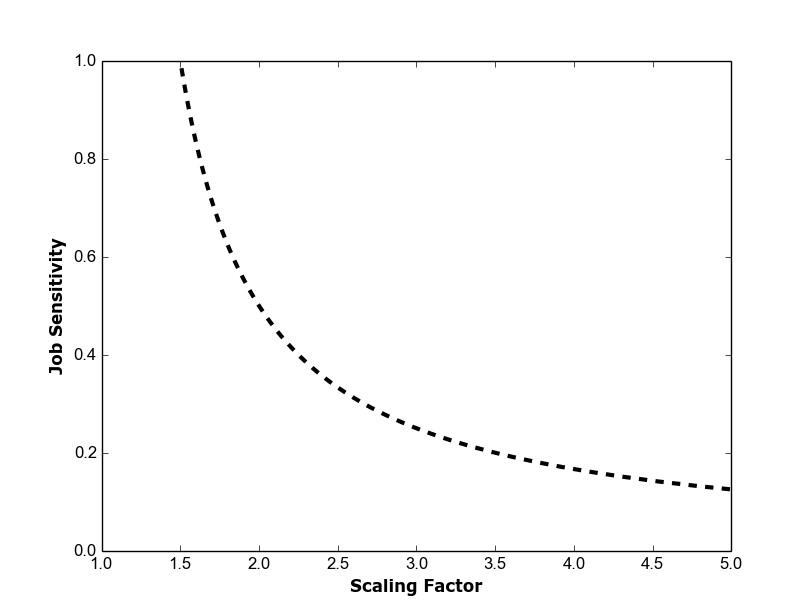}
\caption{Scaling Factor VS Job Sensitivity for acceptable TtS increases}
\label{fig:scalingFactor_VS_jobSens_acceptDeadline}
\end{figure}

This acceptable TtS increase corresponds to the values of scaling factor $\scalingFactor$ and job sensitivity $\jobSensitivity$ displayed in Figure~\ref{fig:scalingFactor_VS_jobSens_acceptDeadline}; 
points below the line correspond to acceptable TtS increase. This information can be used while devising pricing scheme in order to account also for the user satisfaction (for instance, not selecting scaling factor values that would exceedingly lengthen an application).

\subsubsection{\modelOne}
\label{sec:scheme1}

This is the pricing model employed in most HPC facilities. Users pay a price based on the amount of requested resources and the real time-to-solution (wall time) of their job multiplied by the coefficient $\coeffTot$. The total income for the HPC facility is therefore given as the sum of the prices of all jobs that run during the time frame.

In this case (as in the two following ones discussed in Sections~\ref{sec:scheme2} and~\ref{sec:scheme3}) the energy costs are entirely covered by the facility managers, energy savings or increase do not modify the job price for the user which only depends on the TtS. The system owners address this issue by including worst-case estimated energy costs in the cost coefficient $\coeffTot$.

\begin{figure}[bt]

		\subfloat[2d Contour\label{fig:HE_RD_ROI_SYS_sys_gain_norm_2d}]{%
		  \includegraphics[width=.5\textwidth]{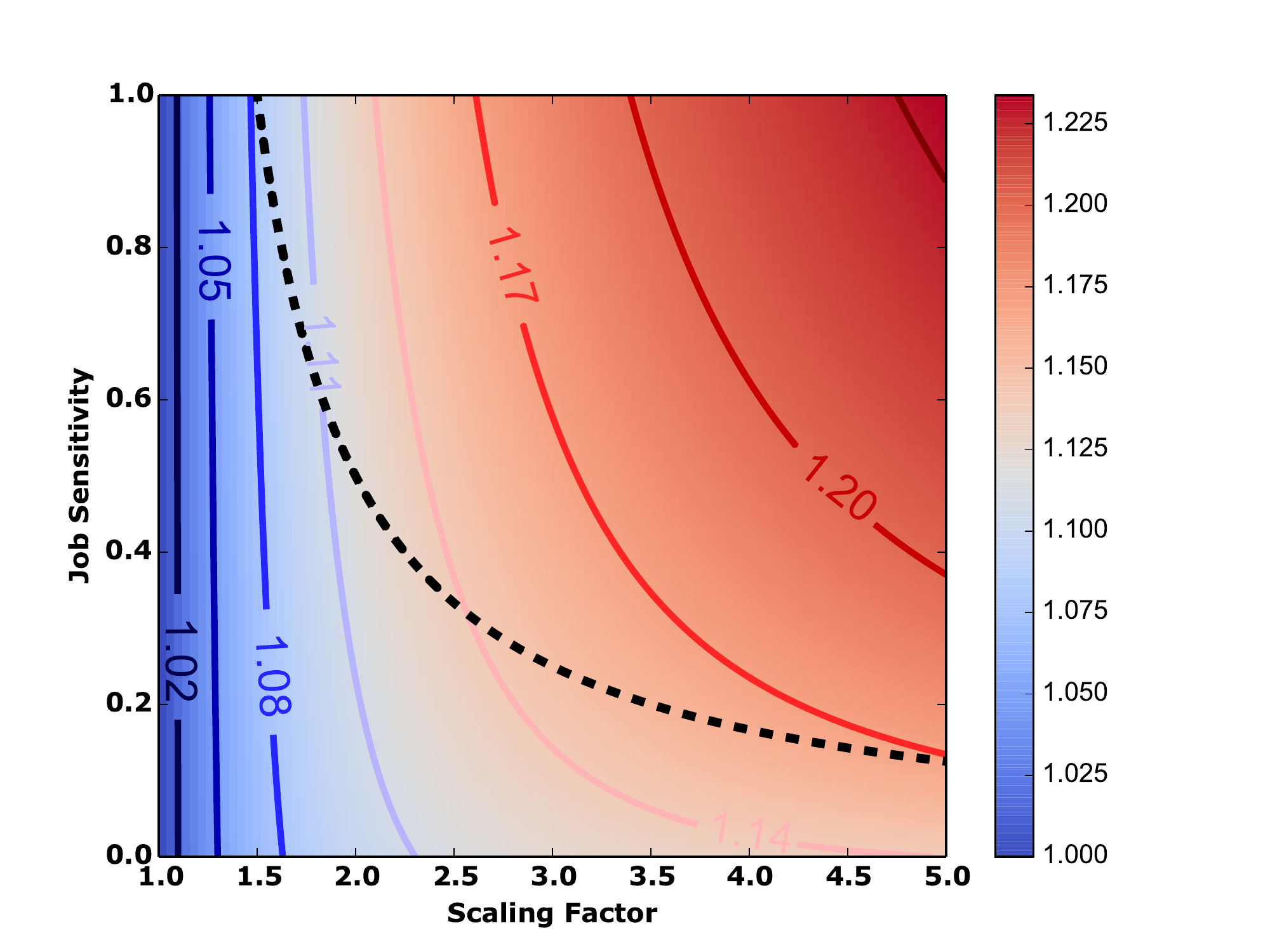}
		}
		\hfill
		\subfloat[3d Surface\label{fig:HE_RD_ROI_SYS_sys_gain_norm_3d}]{%
		\includegraphics[width=.5\textwidth]{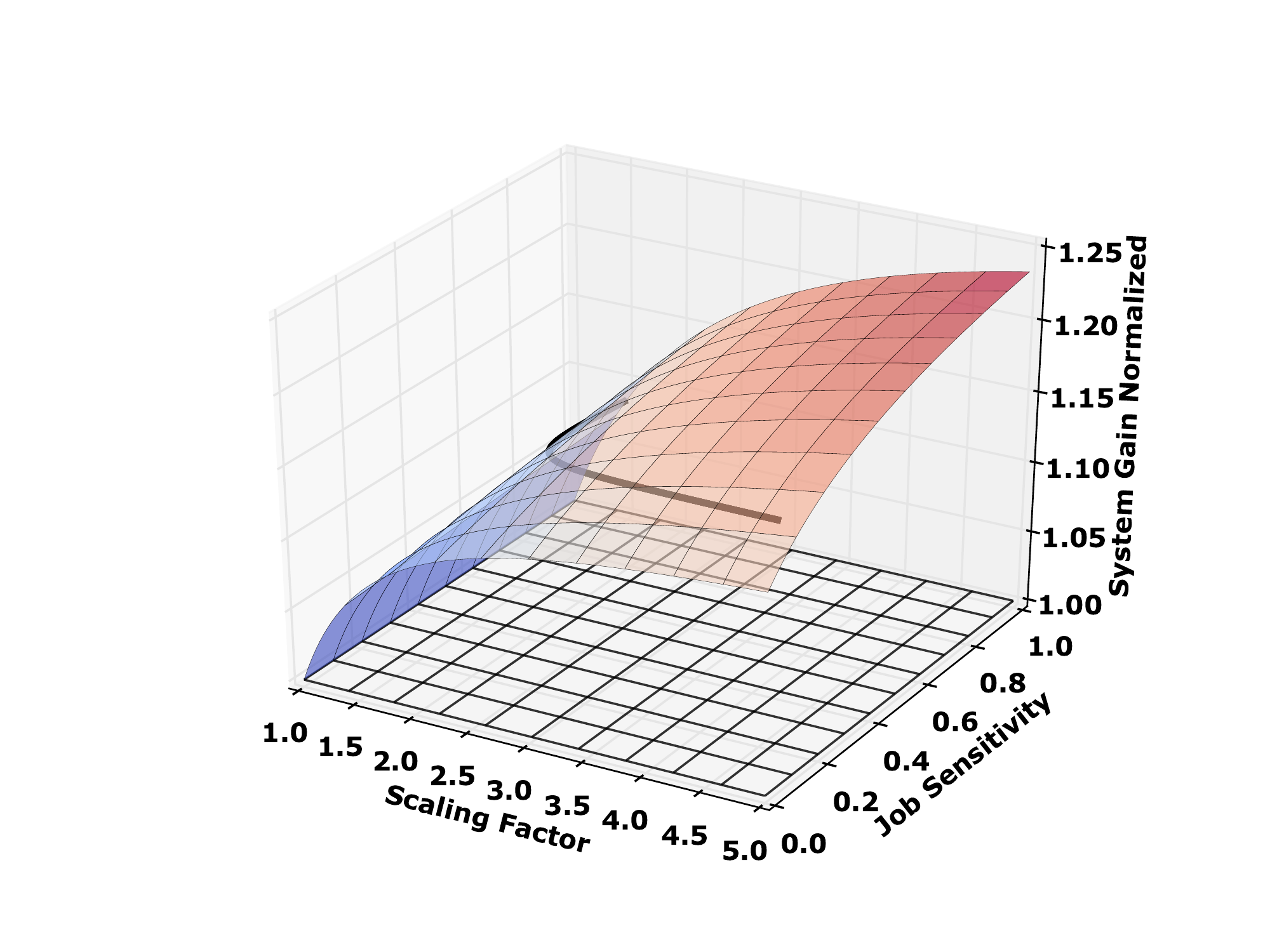}
		}
	
	\caption{\modelOneEM: System Gain Normalized}
	\label{fig:HE_RD_ROI_SYS_sys_gain_norm}
\end{figure}

\begin{figure}[bt]

		\subfloat[2d Contour\label{fig:HE_RD_ROI_SYS_cost_norm_job_2d}]{%
		  \includegraphics[width=.5\textwidth]{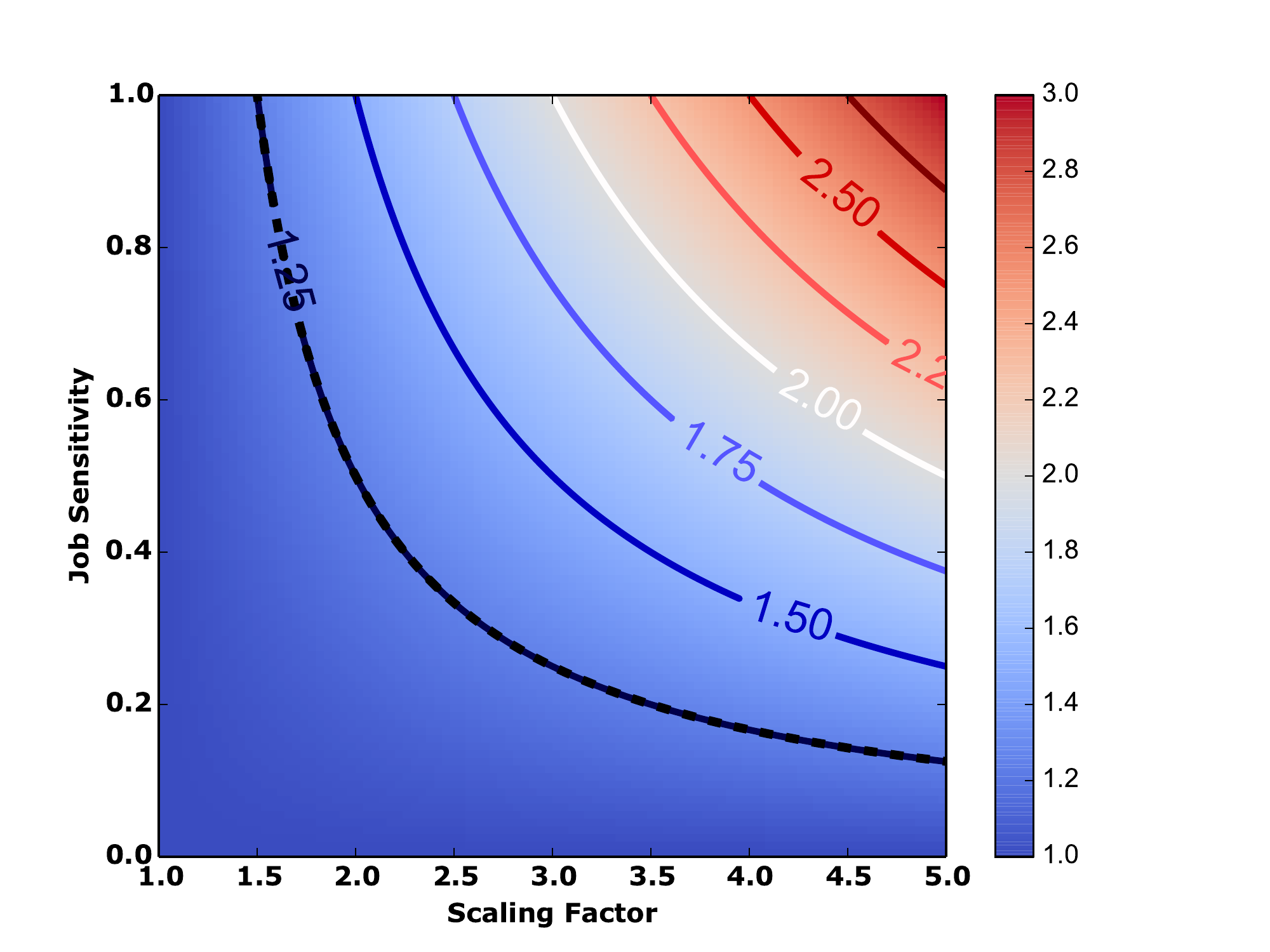}
		}
		\hfill
		\subfloat[3d Surface\label{fig:HE_RD_ROI_SYS_cost_job_norm_3d}]{%
		\includegraphics[width=.5\textwidth]{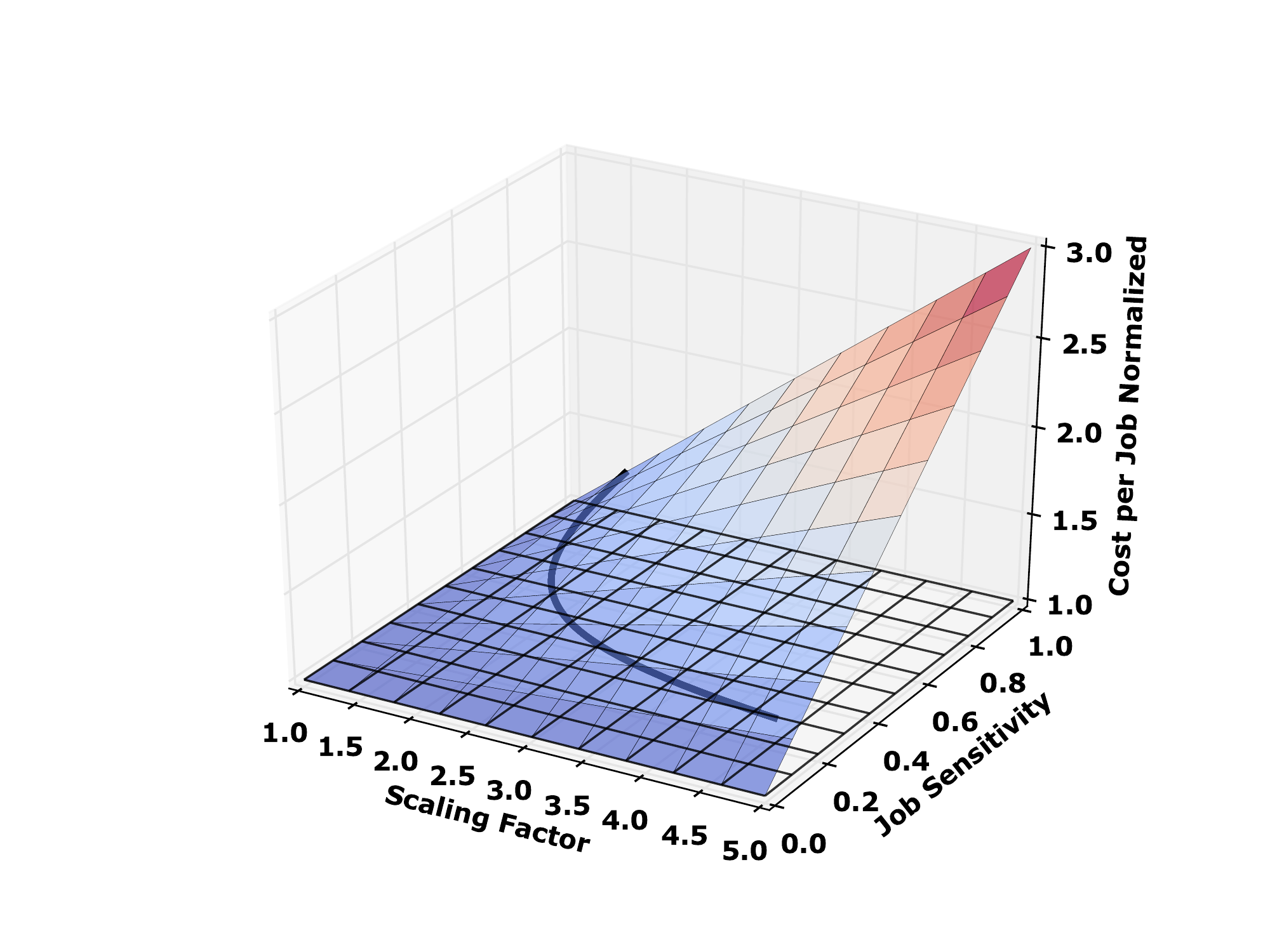}
		}
	
	\caption{\modelOneEM: Job Price Normalized}
	\label{fig:HE_RD_ROI_SYS_cost_job_norm}
\end{figure}

In Figure~\ref{fig:HE_RD_ROI_SYS_sys_gain_norm} we observe the normalized system gain for the \modelOneEM. Fig.~\ref{fig:HE_RD_ROI_SYS_sys_gain_norm_2d} shows in the $x$-axis the scaling factor $\scalingFactor$ and the job sensitivity $\jobSensitivity$ in the $y$-axis; the different colored contours (the lines of points with the same value) indicate the normalized system gain. The same information is presented in three dimensions in Fig.~\ref{fig:HE_RD_ROI_SYS_sys_gain_norm_3d}; here the $x$-axis and $y$-axis indicate again the scaling factor and job sensitivity while the $z$-axis shows the normalized system gain. This kind of coupled plots is used also to look at the normalized job price (Figure~\ref{fig:HE_RD_ROI_SYS_cost_job_norm}) and for the remaining models (see corresponding figures in Sections~\ref{sec:scheme2},~\ref{sec:scheme3}~and~\ref{sec:scheme4}).

The dotted black line plotted in the two-dimensional graphs is the same line seen in seen in Fig.~\ref{fig:scalingFactor_VS_jobSens_acceptDeadline}; combinations of $(\scalingFactor,\jobSensitivity)$ above that line represent conditions where the frequency scaling would make the job TtS longer beyond the point where the user notice the difference (and loss of quality of service -- QoS).

It is quite straightforward to see that with \modelOneEM~the system owner gains more when the scaling factor increases, especially with higher job sensitivity. This happens because the price paid by the users increases due the longer TtS of the jobs. This is clearly shown by Fig.~\ref{fig:HE_RD_ROI_SYS_cost_job_norm}, where the normalized (average) job price rises rapidly together with the scaling factor. If the scaling factor is set to one, the job sensitivity loses its influence and the system gain and job price do not differ from the baseline. This happens with all pricing models. Although this pricing scheme is very enticing from the facility owner point of view, the steep price rises facing the users make its actual implementation almost impossible.

\subsubsection{\modelTwo}
\label{sec:scheme2}
In this strategy the price paid for each job is given by multiplying number of requested cores by the same coefficient of Sec.~\ref{sec:scheme1} and by the job time-to-solution at maximum or nominal frequency. Clearly, the latter quantity can be only known a posteriori or by means of an oracle, a priori. Very precise application and architectural models and monitoring tools could be used to obtain an accurate estimate. The results in this section motivate that this technology would enable power management solutions leading to a win-win situation for the system owner and final users. The income is computed as the sum of all jobs prices. In this case the price per job remains constant, i.e. it is not affected by the reduction in power consumption; for this reason we did not include the corresponding figure. When compared with the default pricing (\modelOneEM) this scheme benefits the supercomputer users while the gains from the system owner's point of view depend on the application scaling factor and job sensitivity. 

\begin{figure}[bt]

		\subfloat[2d Contour\label{fig:HE_FMD_ROI_SYS_sys_gain_norm_2d}]{%
		  \includegraphics[width=.5\textwidth]{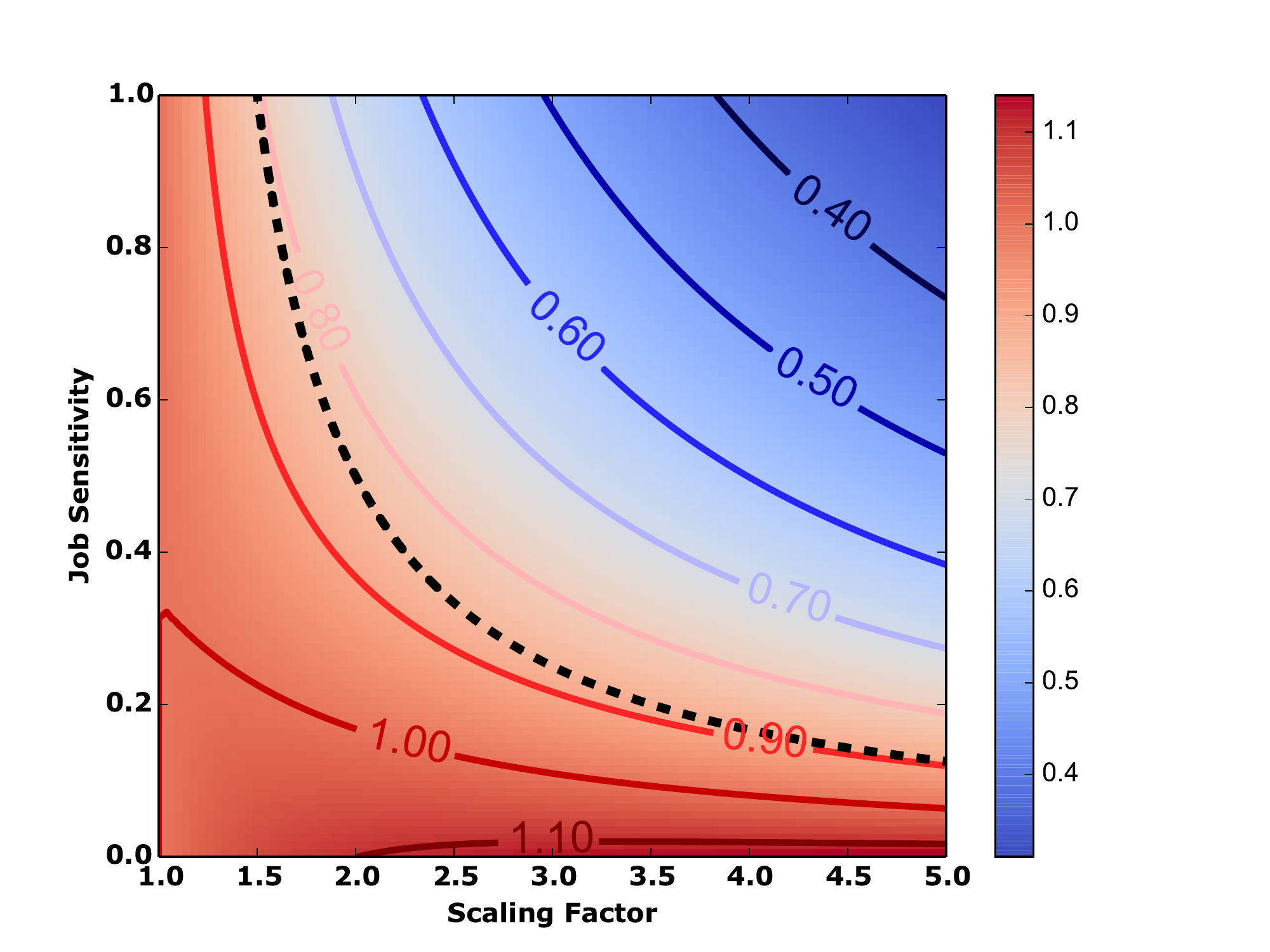}
		}
		\hfill
		\subfloat[3d Surface\label{fig:HE_FMD_ROI_SYS_sys_gain_norm_3d}]{%
		\includegraphics[width=.5\textwidth]{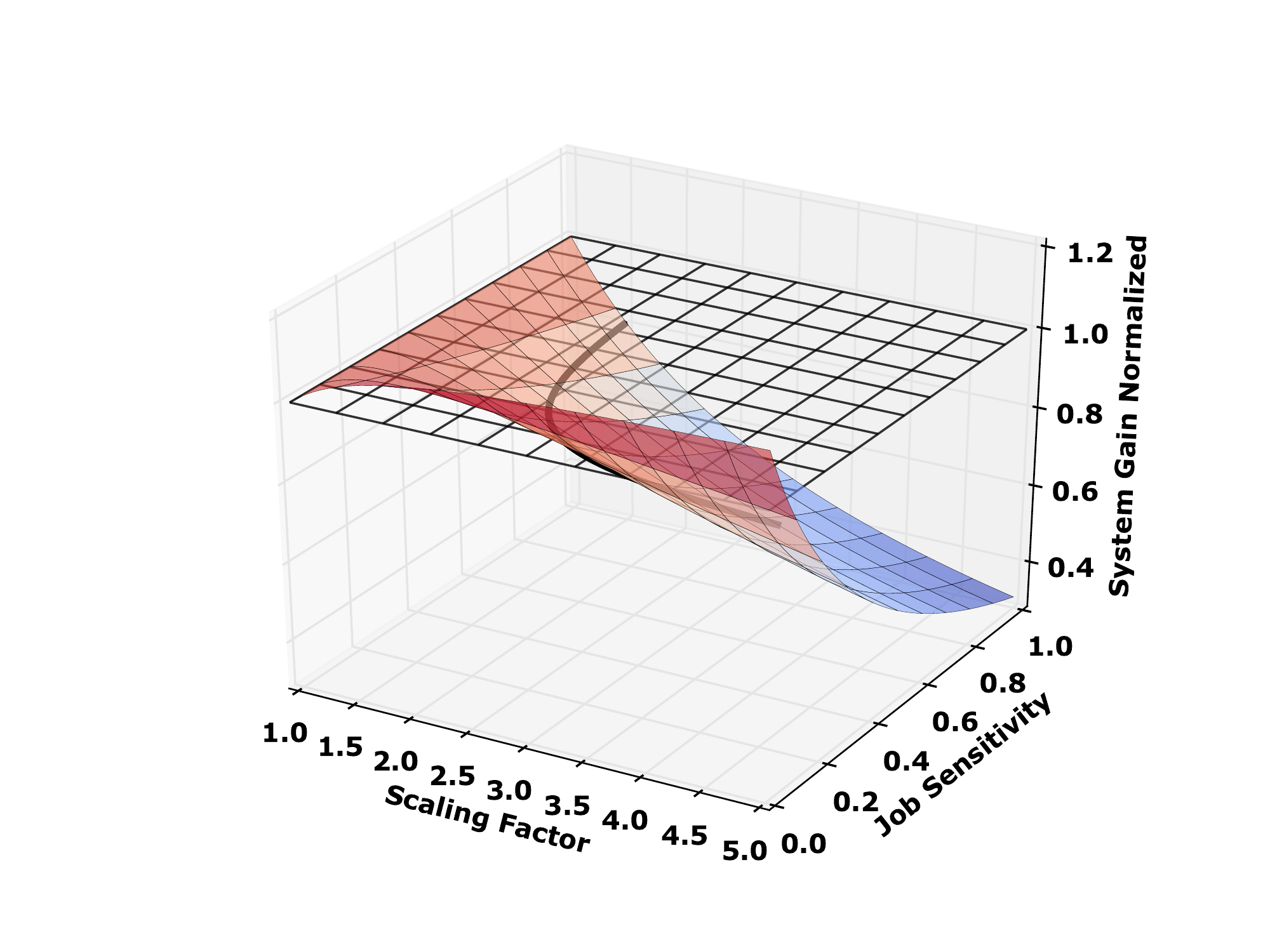}
		}
	
	\caption{\modelTwoEM: System Gain Normalized}
	\label{fig:HE_FMD_ROI_SYS_sys_gain_norm}
\end{figure}

In Figure~\ref{fig:HE_FMD_ROI_SYS_sys_gain_norm} we can observe the normalized system gain for \modelTwoEM. 
As previously noted, with this scheme the price paid by users for each job does not change with the scaling factor because it depends only on the application's estimated TtS while running at maximum frequency. The job price is therefore equal to the baseline one, hence the normalized job price is equal to one in every point. Aside from this relatively trivial consideration, it is worth to note that while the job price remain constant, the system gain drastically changes: when the scaling factor and job sensitivity are relatively low \modelTwoEM~ leads to a larger gain compared to the baseline. This happens because in this case the real job time-to-solution is not too different from the estimated ones and therefore the income loss is lower than the cost saved on energy consumption thanks to the reduced power consumptions. Conversely, when the scaling factor increases the system gain drops since the energy savings does not balance the loss of income relative to the baseline.  

As a final remark, it must be noted from Fig.~\ref{fig:HE_FMD_ROI_SYS_sys_gain_norm_2d} that the area where the system owners achieve a gain (under the red-line with 1.00 marker) is below the user noticeable level (black dashed line). Meaning that the system owner can achieve a gain without inducing QoS loss.  
In this scheme it is essential for the system owner to identify the area delimited by combinations of application sensitivity ($\jobSensitivity$) and scaling factor ($\scalingFactor$) leading to a gain. The system owner assumes the risks for failing it.
To summarize, the actual implementation of this price scheme requires the development of tools for identifying job sensitivity and estimating the application time-to-solution at the maximum frequency. 

\subsubsection{\modelThree}
\label{sec:scheme3}
This pricing model closely resembles the one of Sec.~\ref{sec:scheme2} but tries to solve the problem of estimating the jobs duration at maximum frequency by employing the real job TtS at a scaled frequency with scaling factor $\scalingFactor$. This is done taking advantage of the observation that when reducing a processor frequency of a scaling factor $\scalingFactor$, the time-to-solution can increase at maximum of a factor $\scalingFactor$. For this reason the price of jobs with reduced frequency is discounted by the scaling factor ($\frac{(1-\nonSlowPerc)\jobDurSF}{\scalingFactor}$).

\begin{figure}[bt]

		\subfloat[2d Contour\label{fig:HE_RD-S_ROI_SYS_sys_gain_norm_2d}]{%
		  \includegraphics[width=.5\textwidth]{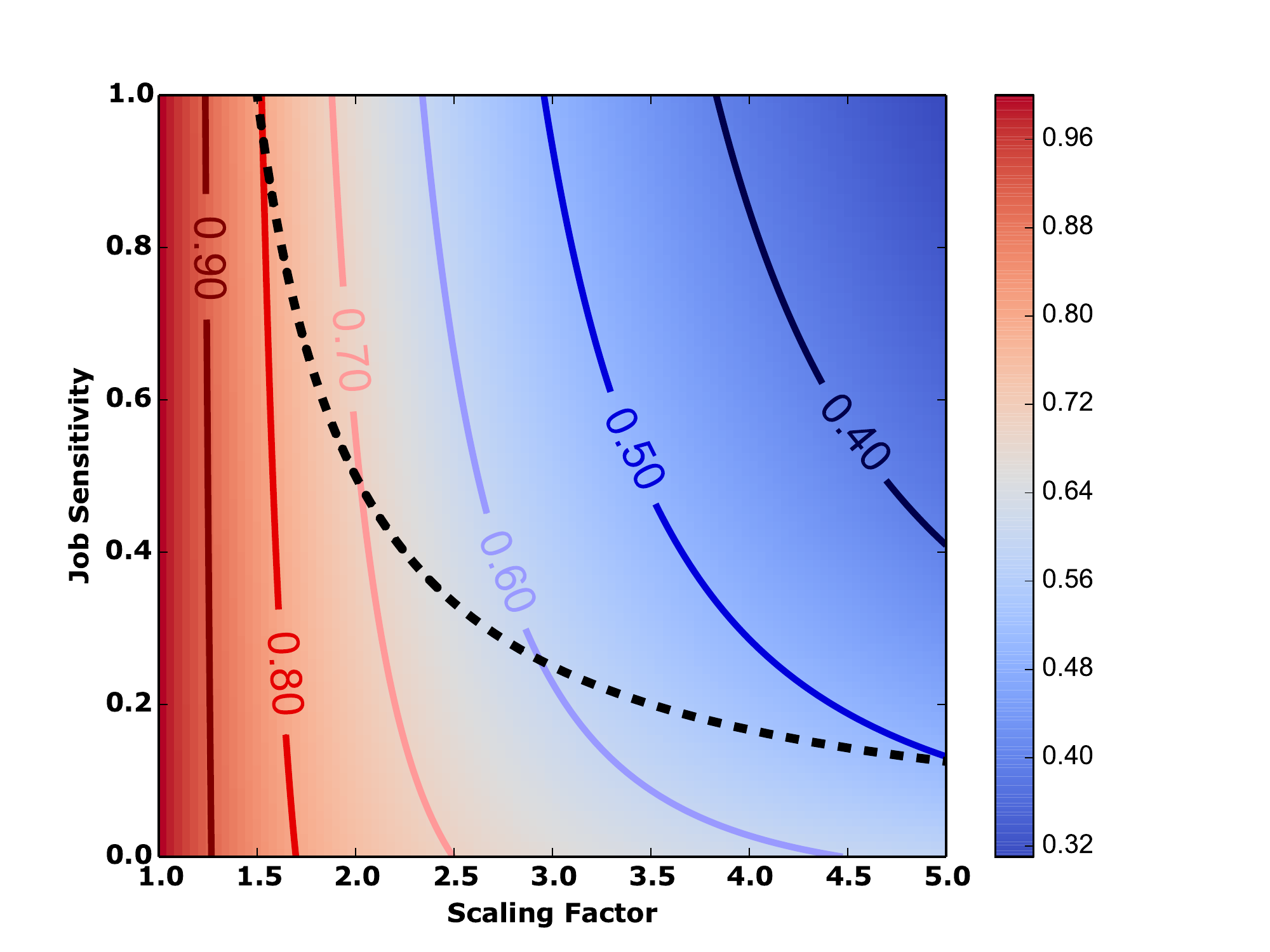}
		}
		\hfill
		\subfloat[3d Surface\label{fig:HE_RD-S_ROI_SYS_sys_gain_norm_3d}]{%
		\includegraphics[width=.5\textwidth]{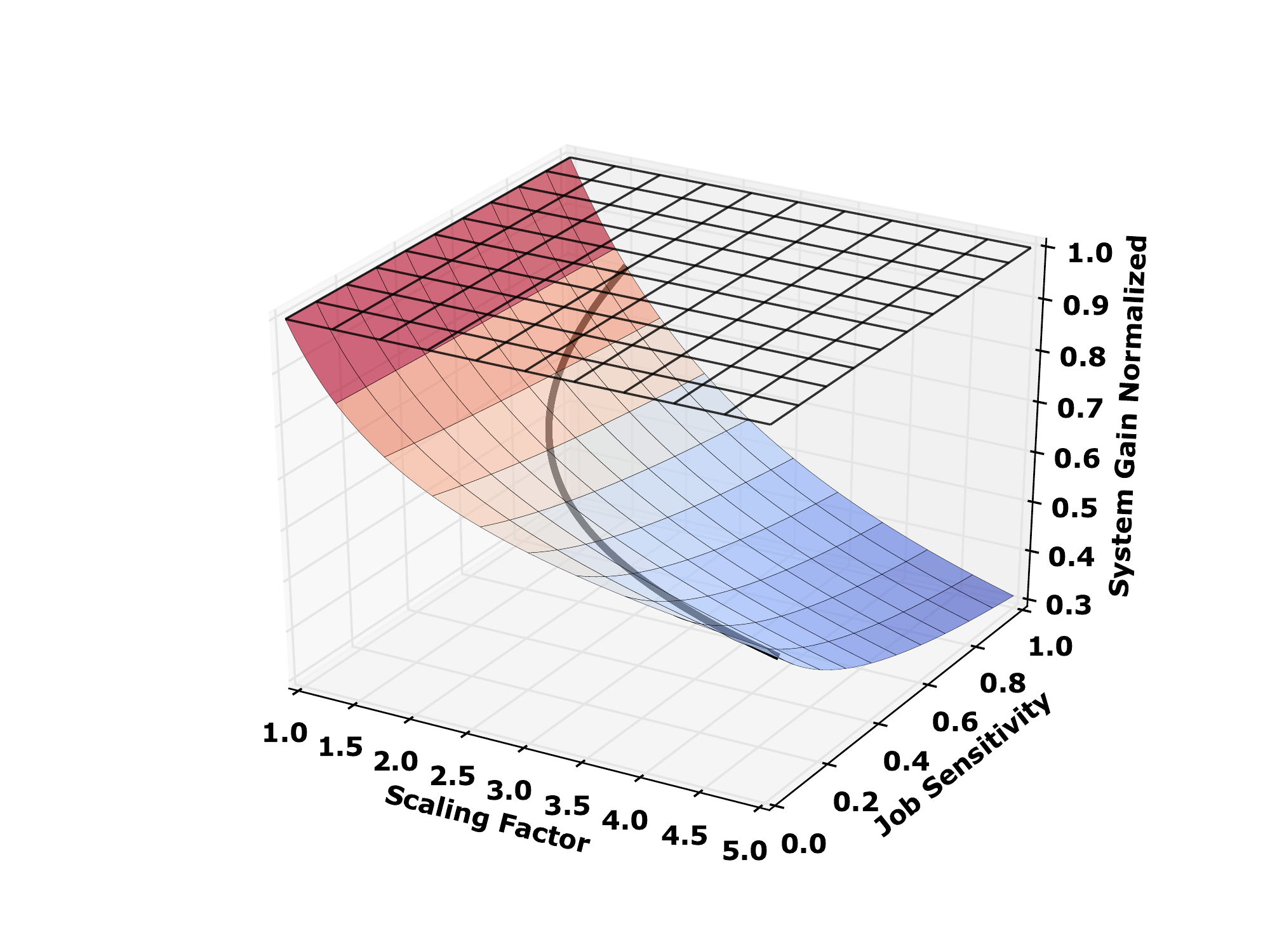}
		}
	
	\caption{\modelThreeEM: System Gain Normalized}
	\label{fig:HE_RD-S_ROI_SYS_sys_gain_norm}
\end{figure}

\begin{figure}[bt]

		\subfloat[2d Contour\label{fig:HE_RD-S_ROI_SYS_cost_norm_job_2d}]{%
	        \includegraphics[width=.5\textwidth]{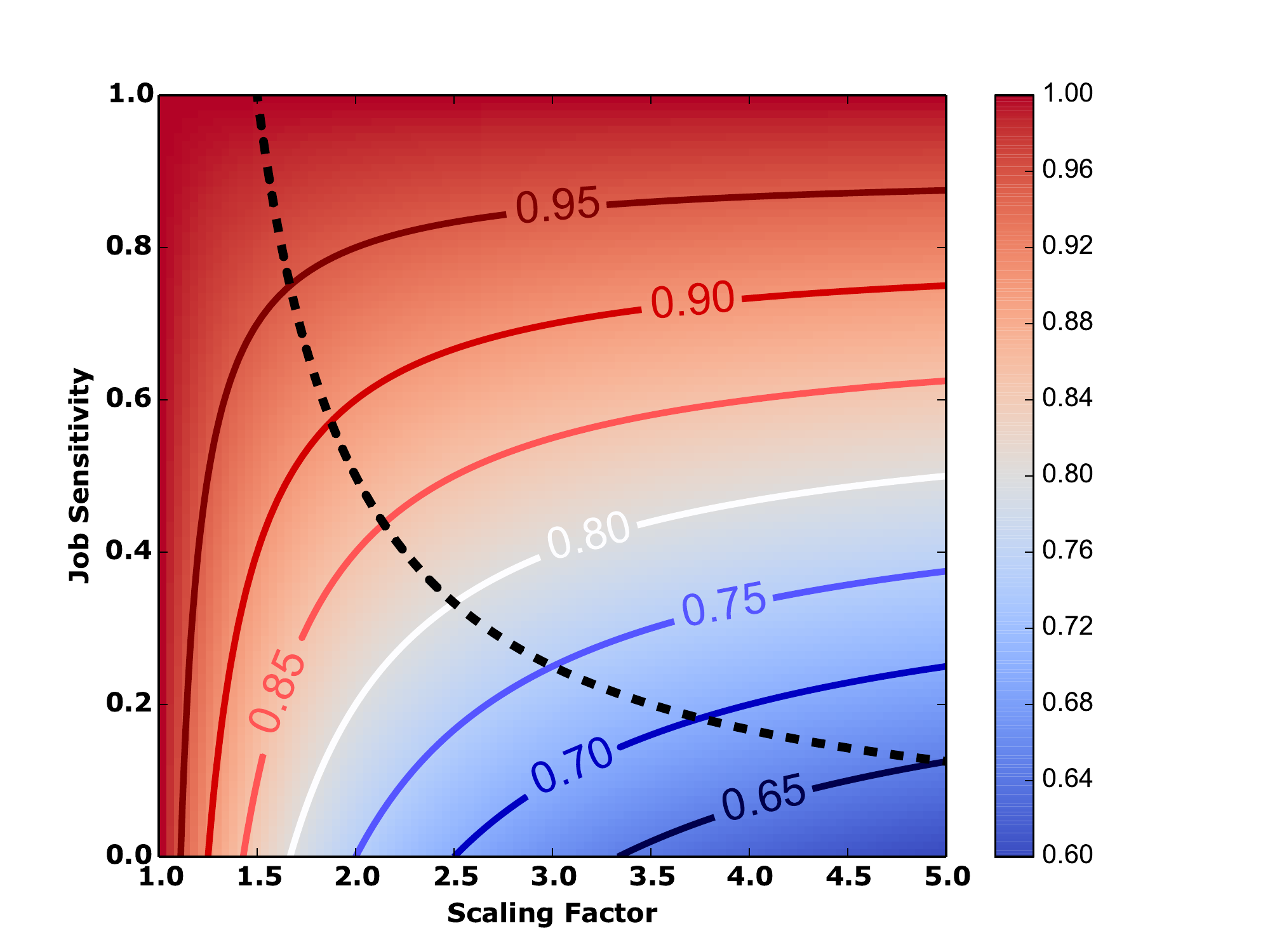}
		}
		\hfill
		\subfloat[3d Surface\label{fig:HE_RD-S_ROI_SYS_cost_job_norm_3d}]{%
		\includegraphics[width=.5\textwidth]{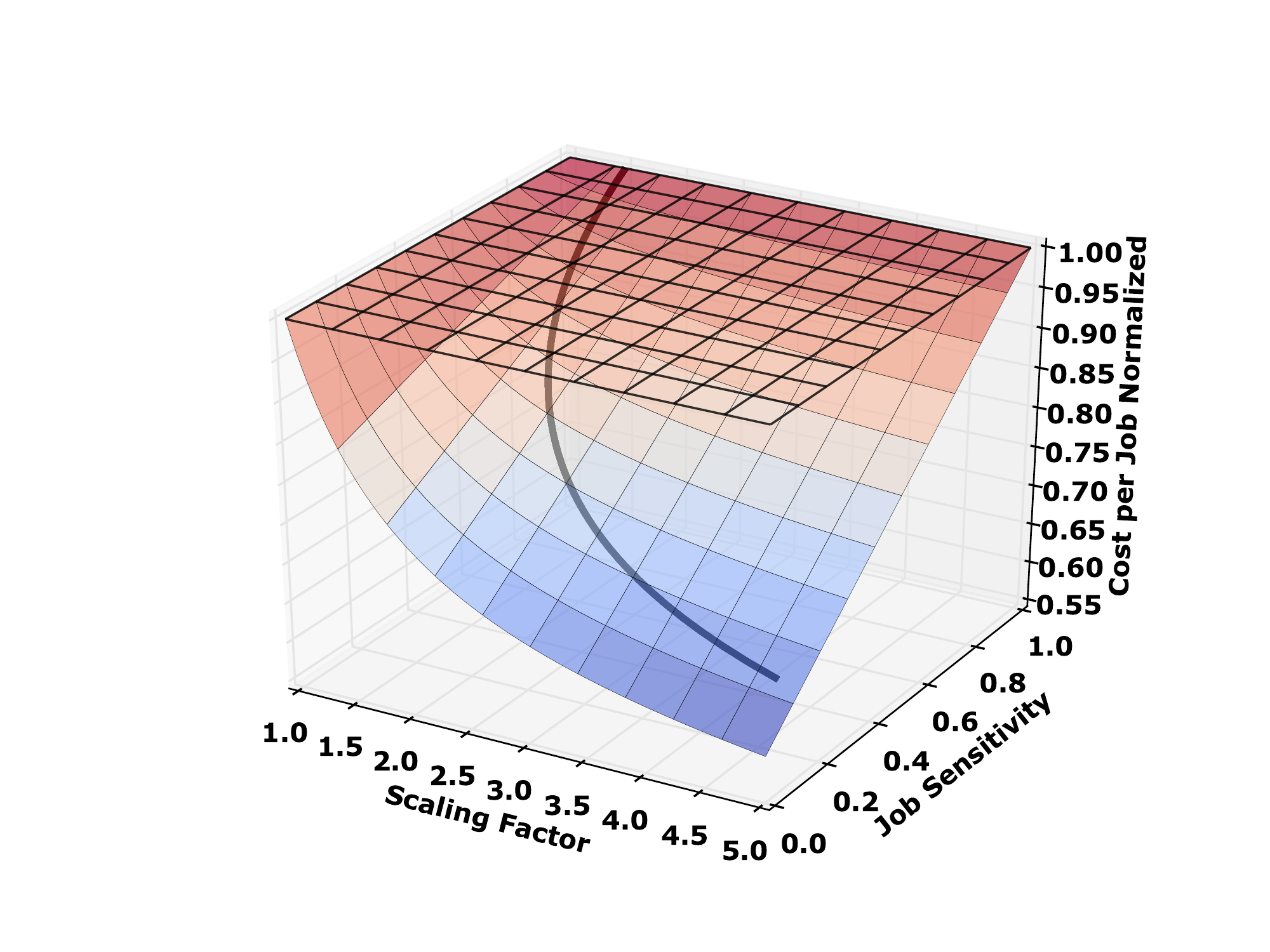}
		}
	
	\caption{\modelThreeEM: Job Price Normalized}
	\label{fig:HE_RD-S_ROI_SYS_cost_job_norm}
\end{figure}

From Figure~\ref{fig:HE_RD-S_ROI_SYS_cost_job_norm} we can notice that this approach is highly favourable from the users point of view, since it leads to markedly diminishing cost when the scaling factor and the job sensitivity increase. The smaller average job price is due to the division by the scaling factor applied to the price of the slowed down jobs. However, for the considered system configuration, this causes a lower system gain w.r.t. the baseline (\modelOneEM) since the energy-related savings are much smaller than the decrease of revenues (see Fig.~\ref{fig:HE_RD-S_ROI_SYS_sys_gain_norm}).

\subsubsection{\modelFour}
\label{sec:scheme4}
With this last pricing schemes, in opposition to the previous ones, the energy cost is not paid by the system owner but it is directly included in the job price. Also in this case the income is given as the sum of all job prices and now each price is composed by two components. The first one depends on the number of requested cores times the TtS (scaled and not scaled) multiplied by the cost coefficient $\coeffSys$; this coefficient is computed excluding the estimated energy costs -- users would not agree to cover the energy costs \emph{twice}. The second component is the cost of the energy of the job, given as the TtS multiplied by the power consumption times the price of the energy ($\electricityPrice$).

\begin{figure}[bt]
	
		\subfloat[2d Contour\label{fig:EE_RD_ROI_SYS_cost_norm_job_2d}]{%
		  \includegraphics[width=.5\textwidth]{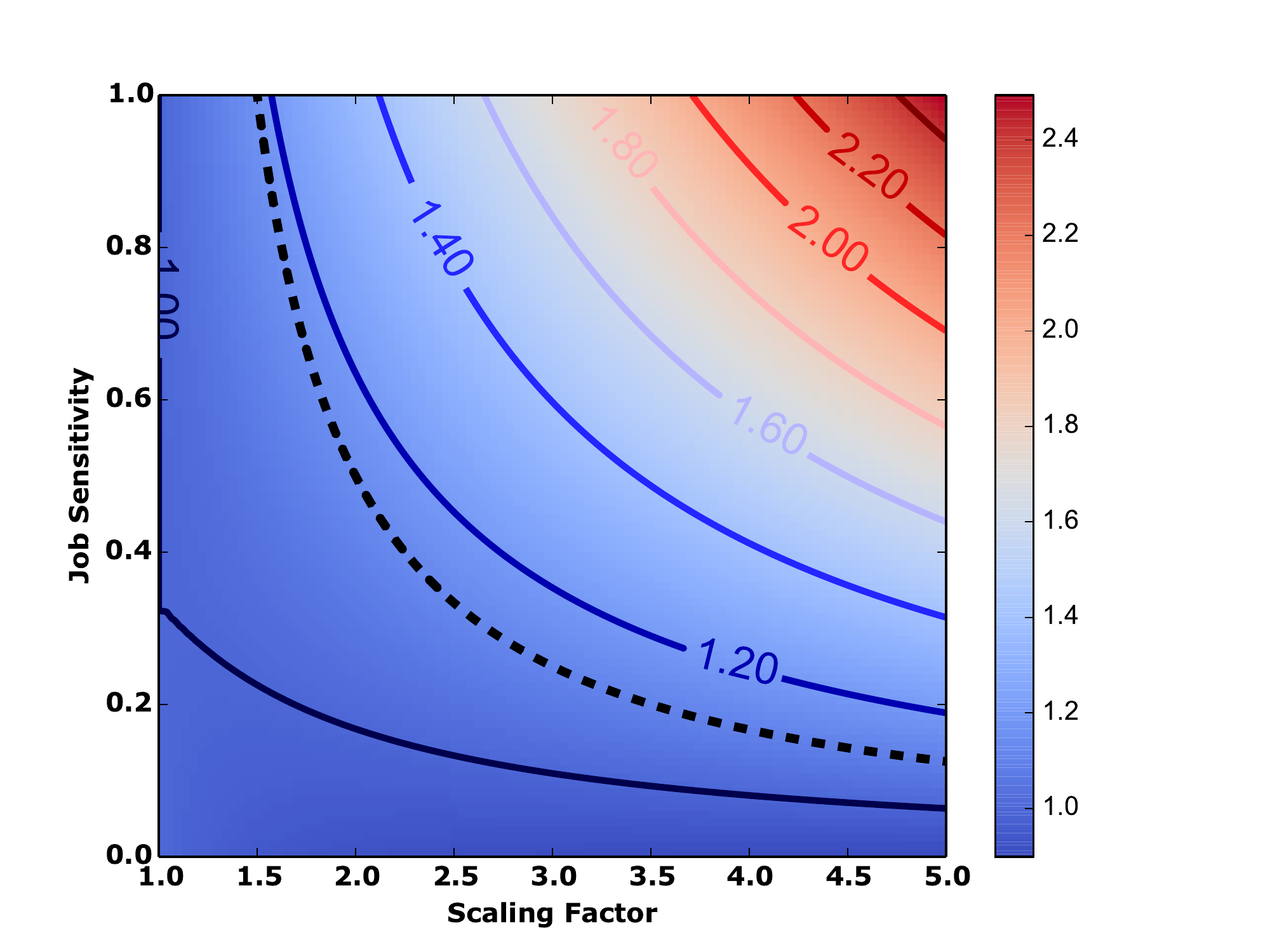}
		}
		\hfill
		\subfloat[3d Surface\label{fig:EE_RD_ROI_SYS_cost_job_norm_3d}]{%
		\includegraphics[width=.5\textwidth]{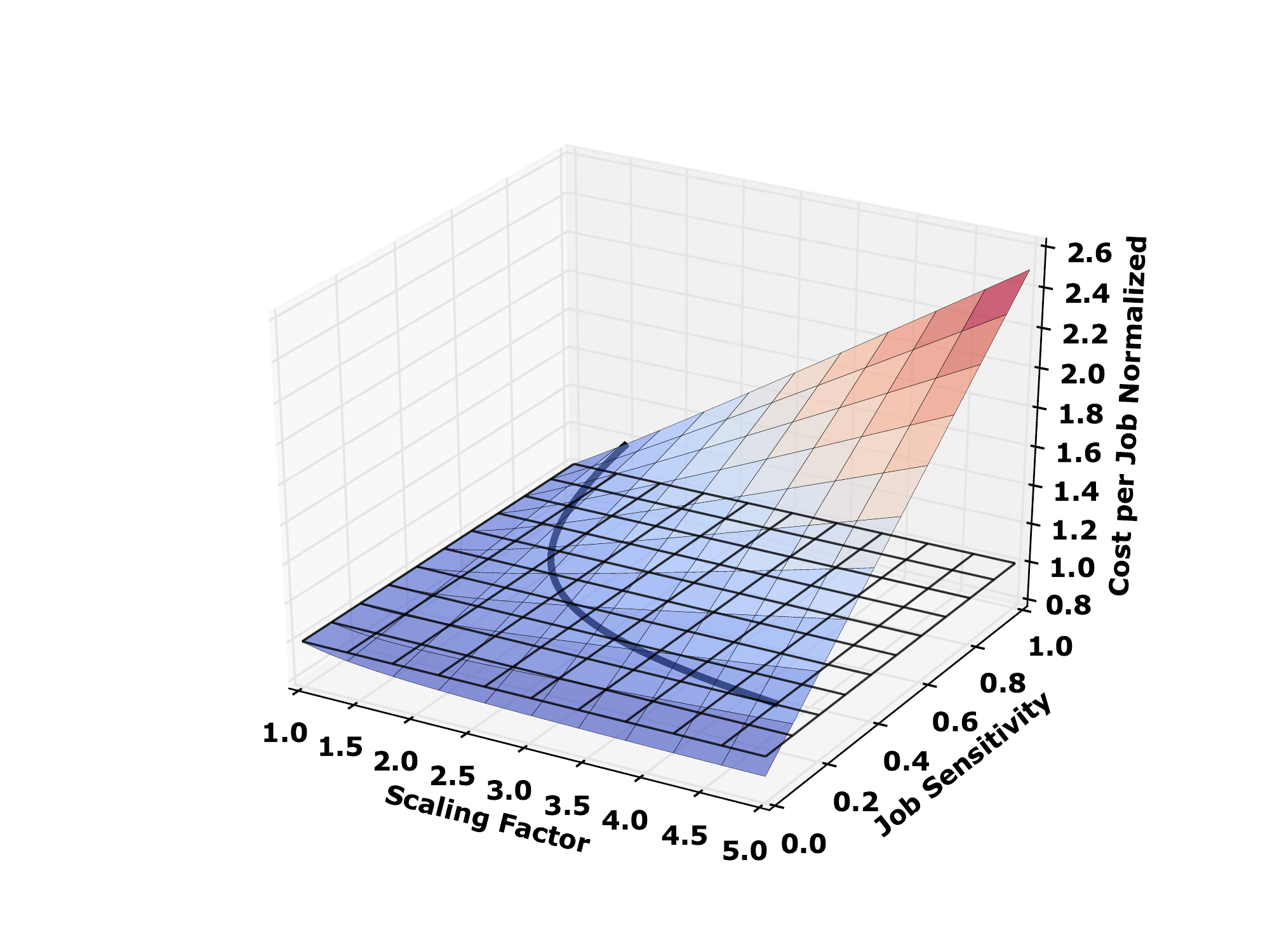}
		}
	
	\caption{\modelFourEM: Job Price Normalized}
	\label{fig:EE_RD_ROI_SYS_cost_job_norm}
\end{figure}

The system gain with \modelFourEM~is constant and therefore also the normalized system gain does not change and it is always equal to the baseline (hence the corresponding figures are not displayed). The possible benefits deriving from the adoption of this pricing scheme stems from the reduction of average job price, as revealed by Figure~\ref{fig:EE_RD_ROI_SYS_cost_job_norm}. With lower values of scaling factor and job sensitivity the normalized job price is smaller than the baseline; when these parameters start rising, the job price follow them accordingly and therefore it surpasses the baseline.
Differently from \modelTwoEM~, this approach shifts the gains and the risks to the final user. 

It does not require estimating the jobs TtS at maximum frequency but only needs a per job energy accounting system.
Clearly, users would need tools for selecting and applying the right power reduction to their applications.

\subsubsection{Pricing Schemes Comparison}
\label{sec:schemes_cmp}

In Table 5 we can see an example of the results of the pricing schemes. Starting from the previous configuration -- based on \emph{Fermi} -- we modified a subset of the input parameters (idle percentage $\idlePerc$, scaling factor $\scalingFactorFixed$ and job sensitivity $\jobSensitivity$); we also varied the amount of cost (per $\timeframe$) due to system depreciation -- expressed as percentage of the total cost. As output we present the difference w.r.t. the baseline, showing both system owner gain and price paid by users, for each pricing scheme. The values in bold highlight the pricing schemes that, under the given condition, manage to bring benefits for both owners and users. 
From the point of view of the system owner positive values are preferable (increased gain), while users prefer negative values (price decrease).

Considering a set point resembling a memory bound application (TtS increase of 20\% as effect of a 2x in frequency reduction) we notice that: 1) \modelOneEM~increases the system gain but penalizes the final user; 2) \modelThreeEM~is beneficial for the user (who gets a discount of 20\%) but generates significant revenue loss for the system owner; 3) \modelTwoEM~and~\modelFourEM~instead lead to noticeable saving without harming the counterpart -- favouring, respectively, the facility manager and the final user. 
Lowering the idle power improves the savings of 2/3 while reducing the depreciation cost of 1/3 doubles the revenues and price reductions achievable by power management strategies. This can reach the 10\% of the total revenues in case of low idle power and long machine turnaround.

The challenge in implementing the \modelTwoEM~is the need to predict what would have been the real application TtS if no power management strategy had been applied; \modelFourEM~only requires the support for accurate per job energy accounting. 

\begin{table*}[bt]
\small\sf\centering
\begin{tabular}{clll@{\qquad}cc@{\qquad}cc@{\qquad}cc@{\qquad}cc}
 \toprule
 \multirow{2}{*}{\emph{Depreciation}} & \multirow{2}{*}{$\idlePerc$} & \multirow{2}{*}{$\scalingFactor$} & \multirow{2}{*}{$\jobSensitivity$} & \multicolumn{2}{c}{\emph{\modelOne}} \qquad & \multicolumn{2}{c}{\emph{\modelTwo}} \qquad & \multicolumn{2}{c}{\emph{\modelThree}} \qquad & \multicolumn{2}{c}{\emph{\modelFour}} \\
 & &  & & \emph{Gain} & \emph{Price Dif.}  & \emph{Gain} & \emph{Price Dif.} & \emph{Gain} & \emph{Price Dif.} & \emph{Gain} & \emph{Price Dif.} \\ 
  \midrule
  \multirow{2}{*}{67\%} & 20\% & 2.0 & 0.2 & 16\% & 10\% & \bf{4\%} & \bf{0\%} & -21\% & -20\% & \bf{0\%} & \bf{-4\%} \\
   & 10\% & 2.0 & 0.2 & 19\% & 10\% & \bf{6\%} & \bf{0\%} & -19\% & -20\% & \bf{0\%} & \bf{-5\%} \\
   \cmidrule{2-12}
\multirow{2}{*}{47\%} & 20\% & 2.0 & 0.2 & 27\% & 10\% & \bf{9\%} & \bf{0\%} & -26\% & -20\% & \bf{0\%} & \bf{-6\%} \\
   & 10\% & 2.0 & 0.2 & 30\% & 10\% & \bf{12\%} & \bf{0\%} & -23\% & -20\% & \bf{0\%} & \bf{-8\%} \\
  \bottomrule
\end{tabular}
\caption{Example: Pricing Schemes Results; the \emph{Gain} and \emph{Price Dif.} columns represent, respectively, the system gains and the price difference, expressed as percentage}
\label{tab:price_schemes_example}	
\end{table*}

\section{Future HPC Scenarios}
\label{sec:alternative_scenarios}

So far, we focused on an existing HPC system with its particular parameters. In this section we are going to explore different scenarios that can be envisioned as near-future evolutions of current supercomputers. As we have seen in Section~\ref{sec:model} two of the main factor impacting the costs faced by system owners are idle power aspects hindering the convenience of frequency scaling, namely the non-null percentage of power consumed by computing units in idle state (the idle power consumption remains constant even if the operating frequency is reduced) and the depreciation costs. The depreciation costs is not influenced by the frequency scaling: if the energy savings are not big enough to compensate the lost income the system owner will face an overall loss. In the system considered as a case study for this work the depreciation costs have a notable impact and they correspond to the 67\% of the total per-time frame expenses.
We consider two cases: 1) energy proportional systems (where the idle power consumption is very low) and 2) low depreciation costs. 

Since the behaviours of the pricing schemes \modelOneEM,~\modelTwoEM~and~\modelFourEM~in the new scenarios are not substantially different than those observed in Sec.~\ref{sec:model_pricing_schemes} we concentrate on \modelThreeEM. Now we want explore the design space to understand if under different conditions this scheme can generate profit also for the system owners; as we have seen before this is the best scheme from the user point of view because it lowers the price paid per job. In the following sections we are going to evaluate the economic viability of \modelThreeEM~in the case of alternative HPC systems, with low idle power consumption (\ref{sec:alternative_scenarios_lowIdle}) and low depreciation costs (\ref{sec:alternative_scenarios_lowMortgage}). 

\subsection{Energy Proportional Systems}
\label{sec:alternative_scenarios_lowIdle}
Several research works have pointed in the direction of energy proportional systems as a possible solution towards improvements in terms of energy efficiency \cite{barroso_energyProp,varsamopoulos2010energy,Lo:2014:TEP:2678373.2665718}. In an energy proportional system the power consumed by its computing nodes scales down proportionally with the load. In our model, this kind of system can be simulated by setting a very low percentage of idle power consumption $\idlePerc$. We analyze the profitability for the system owner using \modelThreeEM; the scheme generates profit if the income for time frame is larger than the expenses (energy costs plus depreciation). We are going to consider the isosurface corresponding to the points where the function $\costTimeframe / \incomeTimeframe$ (total costs divided by income) is equal to $1$. Points \emph{below} the surface represents parameters combinations that are profitable for the system.

\begin{figure}
\centering
\includegraphics[width=.5\textwidth]{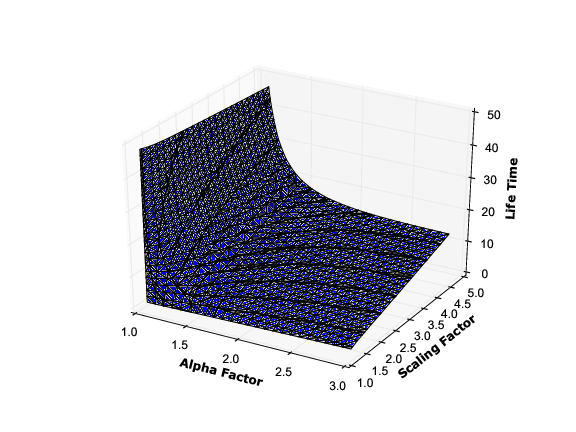}
\caption{System Profitability with low idle power \%}
\label{fig:HE_RD-S_ROI_SYS_sys_profitability_surface_fixedIdle}
\end{figure}

Figure~\ref{fig:HE_RD-S_ROI_SYS_sys_profitability_surface_fixedIdle} considers the system profitability with varying depreciation costs, while maintaining a fixed (very low) value for the idle power percentage ($\idlePerc = 0.01$). In the $x$ and $y$ axis we have the alpha factor $\alphaFactor$ and the scaling factor $\scalingFactor$; the $z$-axis presents instead the system life time $\lifetime$. This parameter is a very good proxy for the depreciation costs impact, since a shorter life time means that the installation costs must be recovered more quickly, hence higher depreciation costs. In the figure, the life time varies in a range of $[1,50]$ years, with a corresponding percentage of depreciation costs (w.r.t. the total time frame costs) of $[88\%,13\%]$. We observe that, with a negligible idle power, the depreciation costs strongly impacts the system gain: with lower life time values is much harder for the system to be  profitable. This happens because if the depreciation costs are the biggest expense source the energy saved through frequency scaling gets negligible while the income loss -- due to dividing the price paid by users by the scaling factor -- becomes preponderant. 

\subsection{Low Depreciation Costs}
\label{sec:alternative_scenarios_lowMortgage}

The second parameter strongly influencing the feasibility of a pricing scheme is the depreciation cost, or more precisely the fraction of the total time frame costs that serve to cover the initial investment expenses. The depreciation costs are regulated by the system installation cost $\systemCostTotal$ and by the expected life time $\lifetime$, that is generally a few years. The continuous quest towards maximum computing performance tends to increase the system installation costs and to squeeze the machines lifetime, but as more nuanced approaches more focused on energy efficiency are gradually taking hold, it is possible to envision slightly different systems where the installation costs decrease and the life time increases. This shift would lead to systems where the depreciation costs impact is less predominant w.r.t. to the energy expenses sustained to operate the machine.

\begin{figure}
\centering
\includegraphics[width=.5\textwidth]{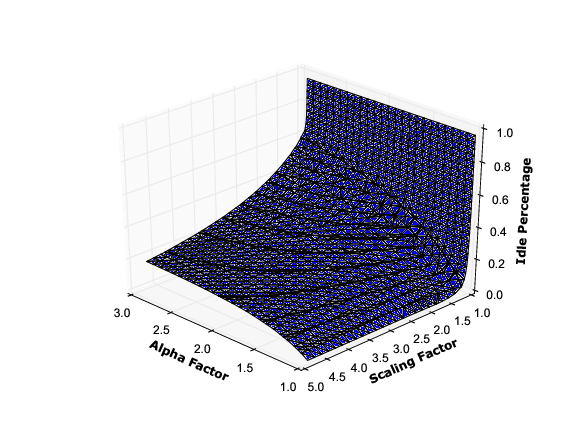}
\caption{System Profitability with low depreciation costs}
\label{fig:HE_RD-S_ROI_SYS_sys_profitability_surface_fixedMortgage}
\end{figure}

In Figure~\ref{fig:HE_RD-S_ROI_SYS_sys_profitability_surface_fixedMortgage} we see the system profitability surface in case of depreciation costs close to zero ($\leq 0.01$\% of the time frame costs). The three axes $x$, $y$ and $z$ represent, respectively, the alpha factor $\alphaFactor$, the scaling factor $\scalingFactor$ and the idle power percentage $\idlePerc$. The points below the surface (i.e. $<2.5, 3.0, 0.2>$) form the region where the system gain is positive (the costs are smaller than the income); as we proceed further from the surface the gain gets higher. With no frequency scaling ($\scalingFactor = 1$) the system is always gaining, due to the remaining model parameters being configured to assure a net profit at maximum frequency (as a baseline). As it was expected, low idle power percentage leads to bigger benefits for the system owner since it allows to consume less power if the frequency is scaled down. We can also notice that higher $\alphaFactor$ values are better for the system owners; this happens because a larger alpha factors means that scaling down the frequency leads to greater energy savings. Finally, we notice an asymptotic behaviour w.r.t. scaling factor: the benefits of decreasing the frequency tend to get thinner and thinner.

\section{Conclusion}
\label{sec:conclusion}
In this paper we tackled the issue of understanding the impact of energy-aware mechanisms in HPC machines. More precisely, we considered frequency scaling as a technique to exchange the power performance of computing nodes in exchange for lower power consumption. Frequency scaling has a clear impact on the energy expenses sustained by a supercomputing facilities and at the same time it strongly influence the accounting mechanism (the price paid by users for using system resources). Our goal was then to provide an instrument capable to analyse the  costs and benefits obtained through frequency scaling in a HPC system. 

We then devised a parametric model inspired by a real supercomputer to simulate the impact of frequency scaling on the system revenue and energy-related costs. We proposed four different pricing schemes and evaluated their effectiveness including the perspectives of both the facility owner and the system users. Our preliminary results indicate that is indeed possible to save energy and curb operational costs via frequency scaling and, at the same time, not to penalize users from a economic point of view. 

As a final takeaway the most valuable strategy to push towards green computing is to shift the cost of the energy consumption to the final user while at the same time providing her instruments for accounting her job energy consumption and scaling the performance level. Letting the system owner play this knob still requires research progress in order to estimate the TtS of applications not perturbed by frequency scaling. In future energy proportional systems, with a longer turn-around, simpler estimation methods will start to pay off as well.

\subsection*{\textbf{Acknowledgements}}
\noindent
  This work was partially supported by the FP7 ERC Advance project MULTITHERMAN (g.a. 291125).
We also want to thank CINECA for granting us the access to their systems.

\bibliographystyle{elsarticle-num} 
\bibliography{bib} 

\end{document}